# Evaporative thermo-fluidics and deposition patterns in surface-active droplets

**Randeep Ravesh**[1,a], **A R Harikrishnan**[2,b], and **Purbarun Dhar**[1,*]

[1] Hydrodynamics and Thermal Multiphysics Lab (HTML), Department of Mechanical Engineering, Indian Institute of Technology Kharagpur, West Bengal 721302, India

[2] Department of Mechanical Engineering, Birla Institute of Technology and Science Pilani, Rajasthan 333031, India

*Corresponding author:* [*] purbarun@mech.iitkgp.ac.in

*E-mail:* [b] ar.harikrishnan@pilani.bits-pilani.ac.in

## ABSTRACT

Droplet evaporation and drying have important implications for coating, inkjet printing, and thermal management applications. Surfactants can modify the drying patterns of particle-laden droplets, but it remains unclear how surfactant molecules themselves modulate the internal hydrodynamics that affect fluid transport and deposition patterns. We investigate the thermo-solutal transport phenomena and deposition patterns during the evaporation of surfactant-laden droplets experimentally and through theoretical scaling-based analysis. It was found that evaporation rates are best explained by considering advection within the droplet, rather than by diffusive transport alone or by changes in interfacial tension. Experiments were conducted using the sessile droplet configuration in the acrylic chamber for both hydrophilic and hydrophobic substrates. The experimental facility was equipped to produce droplets from the dispensing mechanism and capture them under backlight illumination. Infrared thermography and particle image velocimetry (PIV) measurements were conducted during evaporation to illustrate the temperature and velocity distributions, respectively. The constant contact radius (CCR), constant contact angle (CCA) and mixed evaporation regimes were studied using droplet dimensionless geometrical parameters. Experimental observations demonstrated that surfactant molecules modulated evaporation rates, which were further influenced by surfactant concentration and the substrate's wetting state. Sodium dodecyl sulphate (SDS) surfactant molecules enhanced the evaporation rate with an increase in concentration for the hydrophobic surface. In contrast, the evaporation rate increased up to 0.5 CMC and then decreased for droplets on a hydrophilic substrate. The evaporation rates computed from the shadow imaging were explained using the average velocities obtained from the PIV analysis. It was found that advection within the droplet is strongly dependent on surfactant concentration and wettability. Further, the theoretically obtained Marangoni velocities were in close agreement with the experimental values. It was found that Marangoni solutal advection dominates other advection mechanisms, such as Marangoni thermal advection and buoyancy-driven flow. However, surfactant crowding and viscous resistance with increasing surfactant concentration can dampen the increase in solutal advection. The surface tension and viscosity measurements were also conducted with variation in surfactant concentration to understand the suppression of advection by viscous forces. The computation of contact line velocities showed sudden fluctuations, illustrating stick-slip behaviour during droplet drying, complementing microscopic visual observations.

***Keywords***: *Evaporation; surfactants; sessile droplet; PIV; Marangoni advection*

## 1. INTRODUCTION

Evaporation of sessile droplets has remained a domain of major interest within the academic and research community, given its rich physics and extensive implications over a wide range of utilities[1–3]. Over and above this, the evaporation of sessile droplets comprising solvated surfactants, dispersed nanoparticles, or other solutes is a complex transport phenomenon, governed by coupled interfacial and soluto-thermo-hydrodynamic interactions and effects. Over the past decade, extensive experimental, theoretical, and computational studies have been devoted to understanding how surfactants may modify evaporation kinetics, contact line dynamics, internal flows, and final deposition patterns on different substrates.

Early investigations discussed the role of solvated surfactants on evaporation-driven transport processes. Truskett and Stebe [4] showed that even trace amounts of surfactants can alter internal flow fields and particle deposition behaviour significantly by inducing Marangoni stresses during evaporation. Down the timeline, Dugas et al.[5], motivated by micro-fabrication for DNA analysis on chips, showed that surfactant-mediated evaporation dictates solute redistribution and uniformity at small scales. A major theme of such studies is concerned with the surfactant concentration–dependent evaporation rates. Systematic experiments using aqueous surfactant solutions showed non-monotonic trends, where evaporation can be either enhanced or suppressed, depending on surfactant type, concentration, and adsorption kinetics, both at the liquid-vapour and the liquid-solid interfaces[6–8]. Alexandridis et al.[8] showed that structured surfactant solutions exhibit evaporation behaviour distinct from pure water due to interfacial ordering-disordering and resistance to mass-transfer. Similar trends were reported for suspended or pendent droplets[9], and thick liquid-film evaporation[7] of surfactant solutions, highlighting the role of interfacial resistance and altered vapour diffusion pathways in such evaporating droplets.

The nature and role of Marangoni flow modulation within surfactant solution droplets have been explored by research groups. Marin et al.[10] reported surfactant-driven flow transitions in evaporating droplets, where classical thermal Marangoni circulation may be weakened or inverted due to surfactant accumulation and adsorption near the contact line. The findings were supported by theory and/or simulations that dedicatedly incorporated surfactant transport, adsorption–desorption kinetics, and interfacial stresses in the numerical framework[11–13]. Wu et al.[14] reported that soluble surfactants may homogenize surface tension gradients during drying, thereby suppressing Marangoni circulation and altering evaporation dynamics. Comparisons between different contact line models further showed that, in theory, surfactant effects are strongly coupled to wetting assumptions and interfacial boundary conditions[15].

Another crucial aspect for such droplets is the contact line dynamics and dewetting transitions. Zhong and Duan[16] reported surfactant-induced early-stage dewetting transitions, shedding light on the competition interactions between surfactant adsorption and capillary forces during the initial evaporation stages of sessile droplets. Bennacer and Ma[17] extended this study by demonstrating how temperature and surfactant concentration may jointly influence contact line mobility, pinning–depinning cycles, and evaporation regimes. Interestingly, similar aspects were reported for the evaporation of saline droplets on heated substrates, albeit with different contact line behaviour regimes[18,19]. On superhydrophobic surfaces, Aldhaleai and Tsai[20] showed that surfactant concentration governs evaporation mode transitions by modifying both wettability and interfacial mobility, underscoring the combined influence of substrate chemistry and surfactant transport on the evaporative kinetics.

The interplay between surfactants and particles/nanoparticles has received particular attention due to its relevance in coatings, printing, and functionalized fluid samples. Dedicated experiments showed that the interfacial tension dynamics of nanoparticle-surfactant based fluids are starkly distinct, both qualitatively and quantitatively, compared to that of only surfactant solution and only nanoparticle-based-colloids[21,22]. Karapetsas et al.[23] proposed a theoretical framework for droplets laden with both particles and insoluble surfactants, revealing non-trivial and complicated couplings between particle accumulation, surfactant-induced Marangoni stresses, and spatially non-uniform evaporation flux. Experimental studies further showed that particle concentration, size, and morphology can strongly modulate surface tension-driven alterations in contact line behavior, deposition kinetics and patterns, and colloidal self-assembly[24,25], outlining the fact that evaporation kinetics and associated solute-thermo-fluid dynamics in complex fluid sessile droplets cannot be inferred trivially from surfactant or nanoparticle/solute effects alone.

In this context, a series of works by the present authors systematically elucidated the interfacial thermo-fluid-dynamics and evaporation kinetics of nanoparticle–surfactant complex fluid droplets. They showed quasi-universal wetting behaviour of sessile droplets induced by surfactant-capped nanoparticles[26], altered spreading dynamics of sessile droplets due to nanoparticle–surfactant interactions[27], and strong correlations between contact line capillarity and dynamic contact angle hysteresis[28] during droplet-substrate interactions. Also, they reported oscillatory solute-thermal convection during evaporation of pendant droplets of surfactant-nanoparticle complex fluids, revealing a coupled behaviour between evaporation, internal convection, and interfacial chemistry[29] in such complex fluids. These studies emphasized that droplet evaporation of surface-active complex fluids is governed not only by hydrodynamics, but also by thermodynamic and chemical equilibria of solvated species at the interface[30,31].

From application perspectives, biological and living interfaces provide an important motivation for understanding surfactant-solution sessile droplet evaporation. Experiments on plant leaves demonstrated that surfactants significantly affect droplet spreading, coverage area, and evaporation time of liquid pesticides and fertilizers, through interactions with surface roughness, cuticular chemistry, and leaf morphology[32,33]. These reports underscore the practical importance of controlling surfactant-mediated evaporation in spray deposition, agro-chemical delivery, and foliar uptake. Recent efforts have been extended toward multicomponent droplets and surfactant mixtures. Li et al.[34] reported evaporation-induced crystallization of surfactants in sessile multicomponent droplets, revealing complicated phase transitions during drying. Kotsi et al.[35] showed that surfactant mixtures on slightly hydrophobic substrates can exhibit synergistic or competitive effects on evaporation rate and contact line dynamics, depending on the surfactant-substrate interactions.

Despite many studies, a major field of research on the evaporation kinetics of surfactant-solution sessile droplets remains open. The advective pattern within the droplet is affected by various transport mechanisms, including buoyancy-driven flows, surface-tension-gradient-driven flows, capillary flows, and shear arising from the vapour diffusion layer surrounding the droplet. It is not clear about the nature of the dominant transport mechanism impacting the evaporation and deposition patterns with changes in wetting state and surfactant type. The current work focuses on an experimental and scaling-based analytical approach to elucidate the genesis and impact of internal advection during evaporation and drying behaviour. The important transport mechanisms are identified, and non-dimensional numbers are computed to comprehend the complex physics involved in the droplet evaporation. The droplet shadow imaging, infrared thermography, and particle image velocimetry were used as the main diagnostic tools. Both anionic and cationic surfactants are used to conduct the experiments with variation in the wettability. The influence of surfactant concentration relative to the CMC on evaporation rates has been explained by considering both changes in advective pattern and thermophysical properties. The proposed mathematical formulism was found to be effective in predicting the experimentally observed velocities and in deciphering the dominant transport mechanism.

## 2. MATERIALS AND METHODOLOGIES

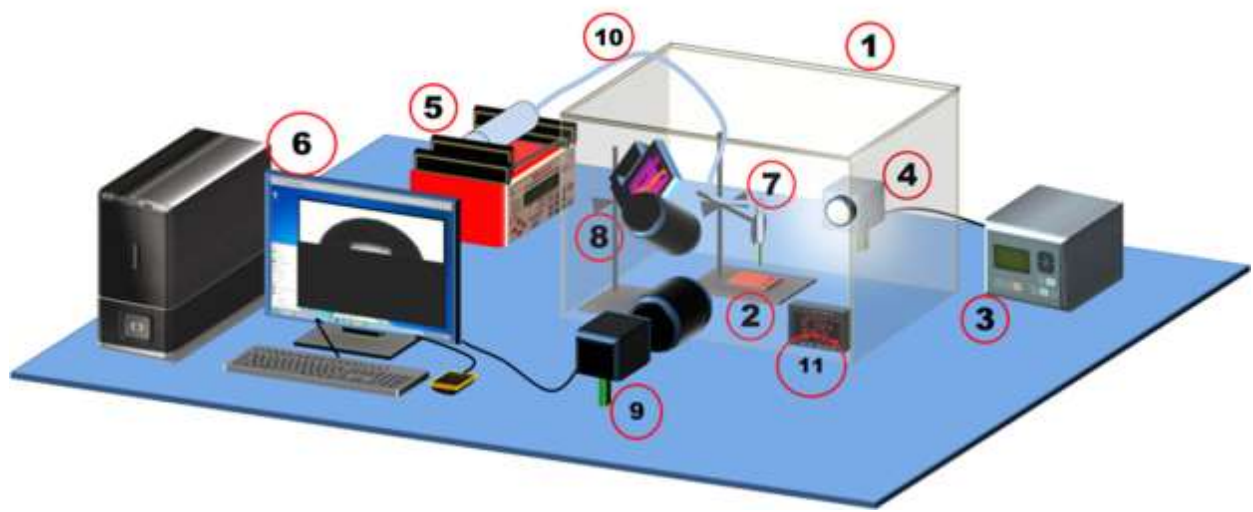

**Figure 1:** Graphic illustration of the experimental setup: 1) acrylic enclosure, 2) substrate with sessile droplet, 3) backlight illumination control unit, 4) light with diffuser arrangement, 5) syringe pump and controller unit, 6) data acquisition computer, 7) needle assembly to dispense droplet, 8) Infrared camera (within the enclosure), 9) CCD camera with microscope probe lens arrangement, 10) drip-tube, and 11) temperature and humidity sensor (within the enclosure). The whole assembly is mounted on a vibration free tabletop. The laser module and sheet optics assembly for the velocimetry studies are not shown here.

### A. Materials

The experiments conducted in the present work utilized cationic and anionic surfactants. The deioninzed water (millipore) of conductivity 1-3 µS/cm was used to prepare surfactant solutions. Sodium dodecyl sulphate (SDS) of purity 99%, Sisco Research Labs, India and Cetylmethyl ammonium bromide of purity 99%, Sisco Research Labs, India, were used as anionic and cationic surfactants, respectively. The experiments were conducted on both hydrophilic and hydrophobic substrates. The glass slides, after cleaning with DI water and acetone, were used as the hydrophilic substrates (ph). The hydrophobic substrate (SHS) was prepared by spray-coating the superhydrophobic solution from Rust Oleum Industries brands, USA. The contact angles for water for hydrophilic and hydrophobic substrates were approximately $40^0$ and $155^0$, respectively. The surface energies are already provided in the earlier work[36].

### B. Experimental setup

Figure 1 shows the experimental setup developed to understand the evaporation of aqueous surfactant solution droplets. The droplet was generated and deposited on the substrate using a glass syringe of capacity 50 µL. The syringe was further connected to the syringe pump and controller unit, thereby forming a droplet-dispensing mechanism. A drip tube was attached from the syringe pump to the dispenser for transporting aqueous surfactant fluid. The images were captured under backlit illumination to calculate droplet geometric features. The high-speed images were captured using a CCD camera (Holmac Optomechantronics, India) fitted with a microscopic probe lens arrangement. The droplet dispensing system, along with the substrate, was enclosed in an acrylic chamber with the temperature & humidity sensors. The entire experimental setup, including the sensors, was mounted on a vibration-free tabletop.

The evaporation process leads to thermal gradients within the droplet. Therefore, the droplet temperature distribution was analysed using an infrared camera (FLIR T650sc). The infrared detector has a resolution of 6410×512 pixels. Furthermore, the detectors have an accuracy of ±0.3 °C over the temperature range 0-100 °C. The internal fluid motion impacts the evaporation of surfactant-laden droplets. Therefore, particle image velocimetry (PIV) was conducted to quantify the internal flow pattern within the droplet. The fluorescent particles of 10 µm diameter from Cospheric LLC, USA, were used as seeding particles. A continuous laser (Reither GmbH, Germany) with a wavelength of 532 nm and a peak power of 10 mW was used for PIV illumination. Particle image velocimetry (PIV) was conducted to capture the evaporation process at 10 fps with a resolution of 1280*980 pixels. ImageJ is used to process images and extract useful geometrical information. The droplet volumes, contact angle, and contact radii were calculated using the spherical cap approximation.

## 3. RESULTS AND DISCUSSION
**A. Evaporation kinetics of surfactant laden droplets**

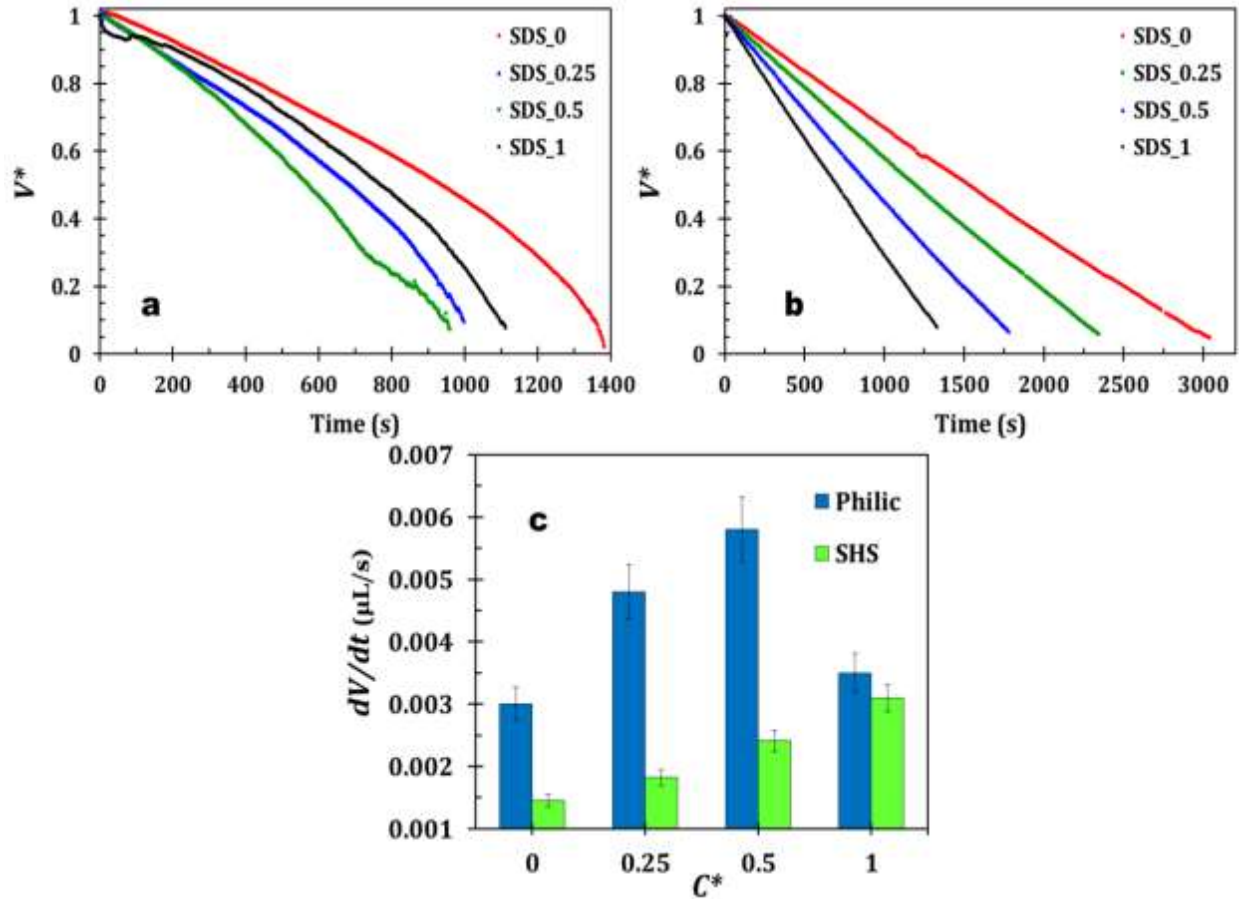

**Figure 2:** a) Non-dimensional droplet volume ($V^*$) with time for SDS solution droplets of different concentrations (expressed as fraction of the CMC) over hydrophilic surface, b) Non-dimensionalization ($V^*$) versus time plots for the same fluids over superhydrophobic surfaces (SHS), and c) temporal rate of droplet volume decay (for the linear regimes in part a and b) as function of SDS concentration for hydrophilic surface and SHS. The SDS_0 symbol denotes that SDS is the surfactant with a concentration of 0 CMC.

The evaporation rates can be understood by analysing the geometrical features of the sessile drop. Therefore, volume ($V$), contact angle ($\theta$), and contact diameter ($D$) are non-dimensionalized to explain the evaporation patterns. Figures 2(a) and Figure 2(b) shows the non-dimensionalized droplet volume ($V^*$) variation with time during the evaporation of a SDS laden droplet for hydrophilic (ph) and superhydrophobic (SHS) cases, respectively.

Moreover, it can be noticed that $V^*$ decreases quasi-monotonically with time (Figures 2a and 2b) for both hydrophilic and superhydrophobic substrates. The droplet decay slope rates and lifetimes rely strongly on SDS concentration and surface chemistry. Therefore, there is coupled influence of surfactant concentration and substrate wettability on the evaporation dynamics of sessile SDS solution droplets. On the hydrophilic substrate (Fig. 2a), increasing SDS concentration up to 0.5 critical micelle concentration (CMC) accelerates volume loss, as reflected in steeper $V^*$ slopes and reduced droplet lifetimes. Figure 2(c) quantitatively compares the linear-regime evaporation rates ($dV/dt$) for both hydrophilic and hydrophobic substrates with variation in SDS concentration. The volumetric decay rate increases from 0.003 µL/s to 0.006 µL/s as concentration changes from 0 to 0.5 critical micelle concentration (CMC). This behaviour is consistent with prior studies showing that sub-CMC concentrations of ionic surfactants enhance evaporation by reducing surface tension gradients and altering internal flow fields[37,38]. At intermediate concentrations, Marangoni stresses induced by surfactant adsorption–desorption kinetics can intensify internal circulation, thereby thinning the thermal and concentration boundary layers at the liquid–vapour interface and increasing vapour flux. However, the trend partially reverses at 1 CMC, where evaporation slows relative to the 0.5 CMC case. Gaalen et al.[13] found that formation of micelles diminished Marangoni convection while Hu & Larson[37] also concluded that surfactant contamination at the interface suppresses Marangoni flow. Therefore, this non-monotonic dependence in the current work aligns with earlier reports that near-CMC surfactant crowding suppresses surface mobility, dampens Marangoni convection, and introduces interfacial resistance to mass transfer.

The effect of surface wettability becomes more pronounced on superhydrophobic substrates (Fig. 2b). Here, droplets exhibit substantially longer lifetimes and more linear $V^*$ decay, indicative of evaporation in a near-constant contact angle mode with reduced contact line pinning. The reduced solid–liquid contact area and enhanced thermal resistance of the trapped air layer are known to suppress heat transfer and evaporation rates[39,40]. Nonetheless, the same non-monotonic influence of SDS concentration persists, suggesting that interfacial transport processes dominate over substrate effects in setting the evaporation kinetics.

For all concentrations, hydrophilic substrates exhibit higher decay rates than SHS, confirming the dominant role of substrate-mediated heat transfer (Figure 2c). Notably, the maximum evaporation rate occurs at 0.5 CMC for a hydrophilic surface, reinforcing the notion of an optimal surfactant concentration where Marangoni-driven enhancement outweighs interfacial immobilization.

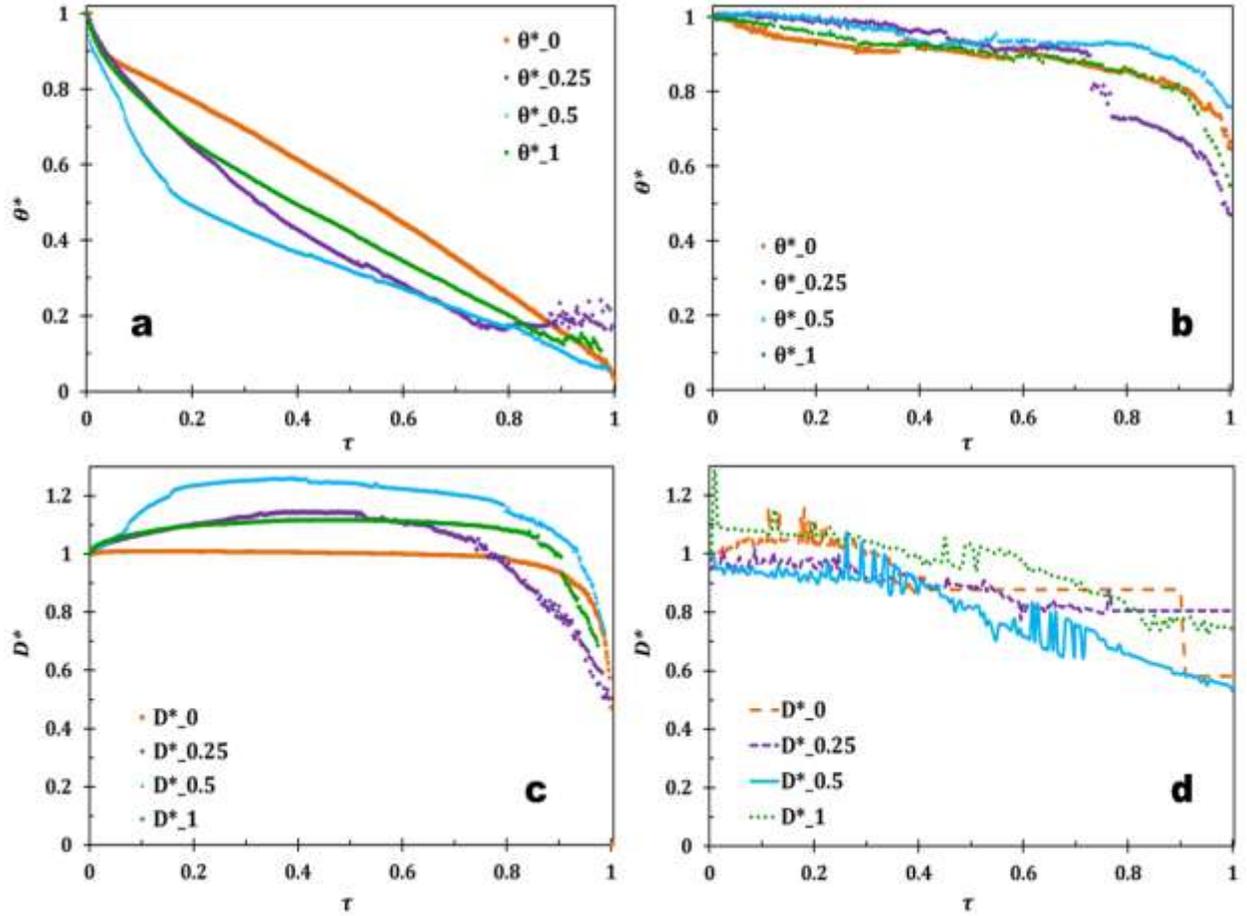

**Figure 3:** Non-dimensional droplet-contact angle ($\theta^*$) vs. time plots for SDS solution droplets of different concentrations (expressed as fraction of the CMC) over a) hydrophilic surface, and b) SHS; and non-dimensional droplet-contact diameter ($D^*$) vs. time plots for SDS solution droplets of different concentrations over c) hydrophilic surface, and d) SHS.

The evaporation dynamics and droplet volume reduction rate can be better understood by analysing the transient variations in the contact angle and contact diameter of SDS solution droplets. Figure 3 (a, b) and Figure 3 (c, d) show variation of non-dimensional contact angle $\left(\theta^* = \frac{\theta}{\theta_i}\right)$ and non-dimensional contact diameter $\left(D* = \frac{D}{D_i}\right)$ with non-dimensionalized time $\left(\tau = \frac{t}{t_f}\right)$, respectively. $\theta_i$, $D_i$ and $t_f$ are the initial contact angle, initial contact diameter, and experimental end time, respectively. The graphs also show the variation in the solution droplets of different concentrations on both the hydrophilic surface and the superhydrophobic surface (SHS). Figure 3(a) shows that the contact angle decreases continuously with time for the hydrophilic surface. The increase in surfactant concentration leads to a steeper decrease in the contact angle up to 0.5 CMC. Further, droplets predominantly evaporate in a pinned or weakly depinned contact-line mode during the linear regime (Figure 3c). Under such conditions, classical diffusion-limited models predict a constant evaporation rate set by the droplet footprint and vapour concentration gradient[41,42]. The observed enhancement in *dV/dt* up to 0.5 CMC (Fig. 2c) indicates a departure from purely diffusion-controlled evaporation, attributable to surfactant-induced Marangoni stresses. At sub-CMC concentrations, non-uniform surfactant distributions along the liquid–air interface generate solutal Marangoni flows that reinforce internal circulation, promoting interfacial renewal and thinning of the thermal and concentration boundary layers. This mechanism effectively increases the local evaporative flux without significantly altering the macroscopic droplet geometry. At 1 CMC, however, the reduction in dV/dt suggests a transition to a surface-mobility-limited regime. Near-saturation of the interface suppresses surface tension gradients and dampens Marangoni convection, consistent with earlier observations of interfacial rigidification and reduced flow intensity in surfactant-laden droplets. The resulting evaporation rate approaches that expected from diffusion-limited scaling alone.

Figure 3(b) depicts the variation of contact angle with non-dimensionalized time for a superhydrophobic substrate (SHS). It demonstrates that evaporation largely takes place in a constant contact angle (CCA) mode. Figure 3(d) shows that the contact diameter fluctuates and the absence of significant pinning of the droplet at all surfactant concentrations on the SHS surface. In the CCA regime, the evaporative flux is reduced due to the smaller solid–liquid contact area and increased thermal resistance associated with the air cushion beneath the droplet. Therefore, longer lifetimes and lower decay rates of droplets across all concentrations are reflected in the evaporation in a near-constant contact angle mode with minimal pinning for SHS surface Nevertheless, the persistence of a concentration-dependent maximum in dV/dt (Fig. 2c) indicates that interfacial transport remains the dominant control parameter, even when substrate effects are weakened. Together, these results suggest that evaporation of surfactant solution droplets is governed by a competition between diffusion-limited vapour transport and Marangoni-enhanced interfacial renewal, with an optimal surfactant concentration near 0.5 CMC. Substrate wettability primarily modulates the baseline evaporation rate by setting the contact-line mode and thermal coupling, while surfactant physics determines deviations from classical scaling.

The evaporation dynamics shown in Fig. 2 can be rationalized within the framework of diffusion-limited vapour transport, with surfactant effects entering primarily through modifications of the interfacial transport prefactor rather than the scaling exponent. For a sessile droplet evaporating quasi-steadily in air, Popov's model[42] predicts a constant volume loss rate $\left(\frac{dV}{dt}\right)$ considering the vapour concentration $C_S$ just above the liquid-air interface, and ambient concentration of water vapour $C_\infty$ as $\sim -D(C_S - C_\infty)f(\theta)R$, where R is the contact radius and f(θ) depend only on the apparent contact angle. The extended linear regimes observed in both Fig. 2(a) and 2(b) are consistent with this prediction, indicating that the dominant evaporation mechanism remains diffusion-controlled over most of the droplet lifetime.

On the hydrophilic surface (Fig. 3a), contact-line pinning maintains an approximately constant footprint, such that variations in dV/dt (Fig. 2c) cannot be attributed to geometric effects alone. Instead, the increase in evaporation rate with SDS concentration up to 0.5 CMC reflects enhanced interfacial renewal driven by solutal Marangoni stresses. At sub-CMC concentrations, non-uniform surfactant adsorption generates surface tension gradients that intensify internal circulation, effectively increasing the vapour concentration gradient at the interface and amplifying the prefactor in the diffusion-limited flux. The subsequent reduction in *dV/dt* at 1 CMC is consistent with interfacial saturation, where surface mobility is suppressed, and Marangoni convection is weakened, restoring evaporation rates closer to classical diffusion-limited values. The effect of cationic surfactant CTAB on evaporation rate is illustrated in Figure A1. The non-dimensional volume droplet volume (V*) variation for CTAB solution droplets of different concentrations over hydrophilic surface, and temporal rate of droplet volume decay as a function of CTAB concentration are illustrated in Figure A1(a) and A1(b), respectively. It can be observed that the maximum evaporation rate occurs at 0.5 CMC, similar to that of the SDS surfactant. However, the evaporation rate remains lower for CTAB, demonstrating that SDS is more effective at increasing the evaporation rate than CTAB. The sharp decrease in contact angle with evaporation is also shown in Figure A2(a). Figure A2(b) demonstrates the non-dimensional wetting diameter with time for evaporating sessile droplets at different CTAB initial concentrations. It can be observed that the wetting diameter remains constant for most of the evaporation, illustrating droplet pinning.

For droplets on superhydrophobic surfaces (Fig. 3b), evaporation occurs predominantly in a constant contact angle mode, resulting in reduced heat transfer through the substrate. These geometric and thermal effects shift the baseline evaporation rate downward across all concentrations. Overall, the data indicate that surfactants modulate evaporation by altering interfacial transport efficiency, while the fundamental diffusion-limited scaling of droplet evaporation is preserved.

## B. Influence of internal hydrodynamics on evaporation

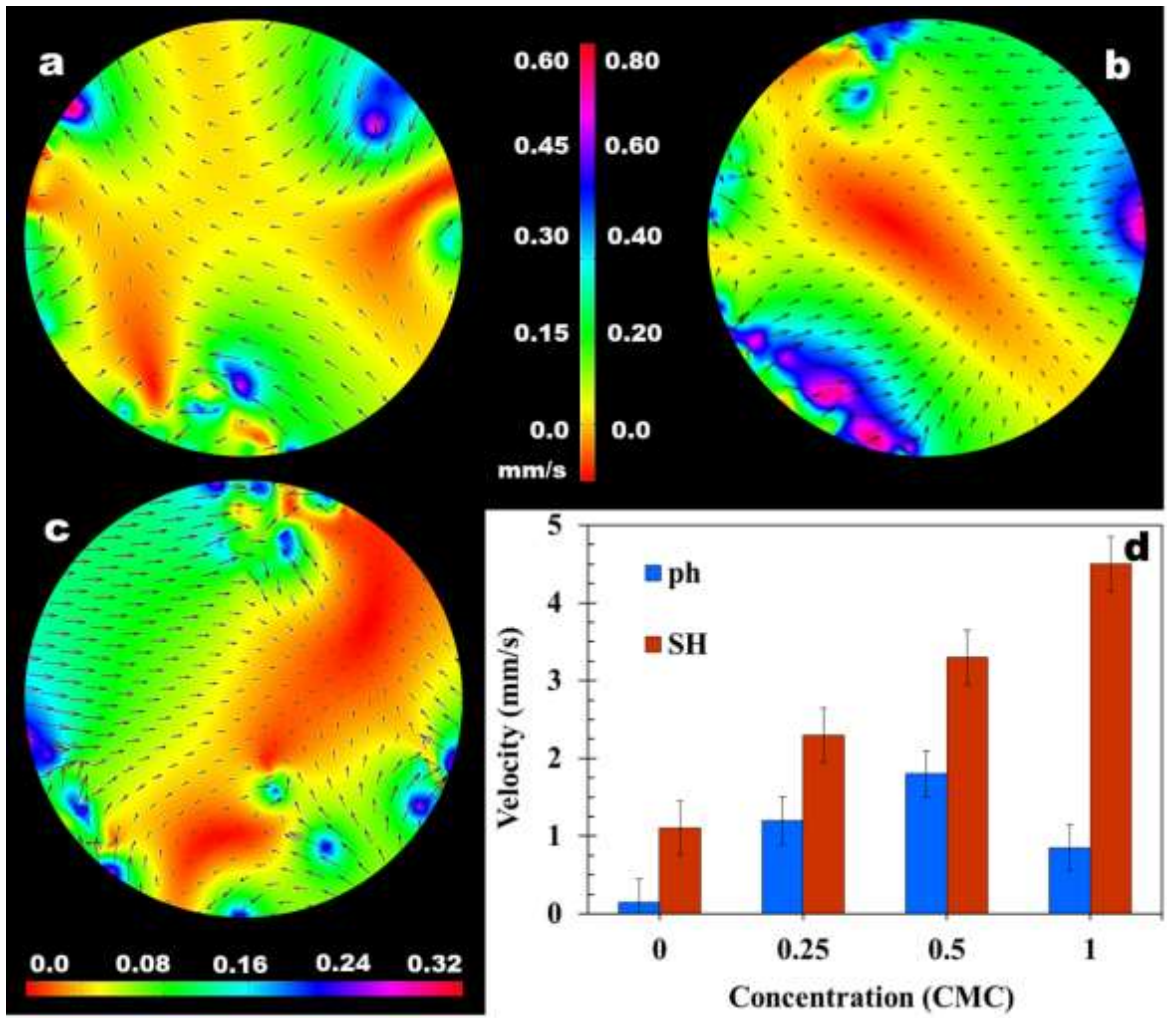

**Figure 4:** Temporally averaged velocity contours (from analyses of PIV experiments data) for SDS solution droplets on hydrophilic surface at: a) 0.25 CMC, b) 0.5 CMC, and c) 1 CMC concentrations. The spatio-temporally averaged velocity values have been shown in part d) for both hydrophilic (ph) and superhydrophobic (SH) cases (see Figure A3 in appendix for the velocity contours on superhydrophobic substrate).

The characterization of the flow field is necessary to explain the dominant mode of advection during droplet evaporation. It is established that the fluid motion inside the droplet is important for explaining the evaporation patterns. The pinning of the droplet causes fluid to move towards the droplet's edges to replace fluid that has evaporated. This is called "capillary flow". The fluid at the edges has a higher surface-to-volume ratio. Therefore, higher evaporation rates occur at the edges. The uneven evaporation across the droplet causes temperature gradients, which in turn lead to surface tension gradients. These surface-tension gradients drive fluid motion from lower to higher surface tension, a process known as thermocapillary advection. Surfactant molecules are also transported towards the droplet's periphery, where they can be deposited and lower the surface tension. These adsorption and desorption phenomena create surface tension gradients and affect wettability. Therefore, differential surfactant arrangements during evaporation create the solutal Marangoni advection. Both solutal and thermal Marangoni effects can complement or compete in their cumulative impact on evaporation rates.

The particle image velocimetry (PIV) was conducted to understand the role of advection generated by surfactant molecules in the droplet. Surfactant molecules affect the advection pattern within the droplet, which in turn influences droplet evaporation rates. Figure 4 shows the temporally averaged velocity contours of droplets evaporating on the hydrophilic substrate with variation in SDS solution concentration. The SDS concentration variation includes (a) 0.25, (b) 0.5, and (c) 1 CMC. The contours shown in Figure 4 are top-view contours of the droplet, as observed through an inverted microscope. The PIV is performed at the horizontal mid-plane for the hydrophilic surface. The region of the droplet near the liquid-gas interface is not shown, and the major central portion of the droplet has been used for the PIV analysis to avoid stray vectors that are common near interfaces of such microfluidic systems. The averaging of the 1000-1200 velocity contours was performed to obtain the final velocity distribution. PIV was conducted in the initial phase of experiment to minimize the effect of change in bulk surfactant concentration within 10% of initial value. Figure 4(d) shows the spatio-temporally averaged velocity values for both hydrophilic and hydrophobic surfaces. It can be observed that velocity increases from ~ 0.25 mm/s to 1.75 mm/s at 0.5 critical micelle concentration (CMC). However, the velocity reduces at 1 CMC for hydrophilic substrate. The velocity values obtained from PIV can be directly related to the evaporation rates shown in Figure 2(c). The evaporation rates were found to increase with surfactant concentration up to 0.5 CMC for the hydrophilic substrate, consistent with the increase in droplet velocity. The surfactant crowding at 1 CMC decreases the velocity and evaporation rate. The increase in soluto-thermal advection leads to higher flow velocities at the surfactant-laden droplet interface, thereby increasing the shear of the vapour layer surrounding the droplet. This "shear-driven flow" assists in recharging the vapour concentration gradient across the diffusion layer.

The PIV is performed at the vertical midplane for sessile droplet evaporation on SHS case. Figure 4(d) shows the dramatic rise in the velocity from ~ 1 mm/s to ~ 4.5 mm/s with an increase in surfactant concentration from 0 to 1 CMC. The temporally observed contours are shown in Figure A3. It should be noted that contours depicted in Figure A3 are side views and regions of droplet near the liquid-gas and liquid-solid interfaces are not depicted. The averaged velocity continues to increase from 0 CMC to 1 CMC in Figure 4(d), which complements the increase in evaporation rates observed in Figure 2c. Therefore, surfactant drives the increase in advection, which in turn generates shear at the interface. The shear leads to increased evaporation of aqueous solution droplet on the superhydrophobic substrate case.

Now, we have appreciated that aqueous surfactant solution droplets enhance advection, but we have avoided discussing the viscous effects of the surfactant molecules. It must be realised that the continuous evaporation of the droplet alters the bulk surfactant concentration. Specifically, evaporation will increase the surfactant concentration in the remaining droplet over time due to water loss. This transient change in surfactant concentration influences thermophysical properties, which in turn affects transport behaviour within the droplet.

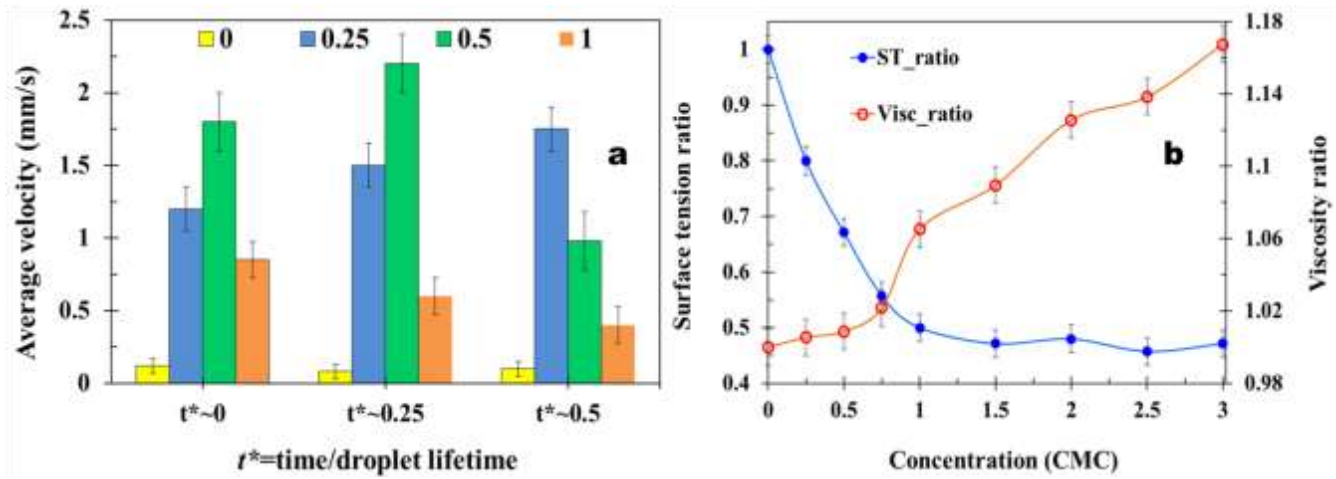

**Figure 5:** (a) Average velocity variation at different dimension times, (b) Variation of surface tension ratio and viscosity ratio with change in critical micelle concentration

Therefore, we measured the viscosity and surface tension with change in surfactant concentration. To explain the effect of transient surfactant concentration on flow velocities inside the droplet, we have also performed PIV at various intervals during evaporation. Figure 6(a) depicts the variation in average velocity at time (t* = instantaneous time/droplet lifetime) ~ 0, 0.25, and 0.5. The velocities are compared to illustrate the effect of SDS surfactant concentration relative to CMC. The average velocity at t*~ 0 is higher for the surfactant-laden droplet compared to pure water. For instance, average velocity increases by 376% from increasing SDS concentration 0 CMC to 0.25 CMC. Moreover, increasing surfactant concentration leads to higher velocities over time. For instance, the velocities at t*~0.25 for concentrations of 0.25 CMC and 0.5 CMC rise by 26.1% and 20.7%, respectively, compared to their values at t*~0. The velocity is enhanced by the positive effect of the surfactant on thermal and solutal Marangoni advection.

Interestingly, the velocity declines by 50% at critical micelle concentration (1 CMC) compared to 0.5 CMC at t*~0. Therefore, velocity does not continuously increase upon the addition of surfactant. Further, the velocity for 1 CMC at t*~ 0.25 reduces by 28.6 % compared to its value at t*~0. It clearly demonstrates the effect of surfactant crowding at the liquid-vapour interface, which ultimately impedes the evaporation. The decrease in velocity indicates that advective transport within the droplet weakens, as water loss causes surfactant molecules to concentrate and increase the surfactant concentration. Moreover, increasing surfactant concentration enhances droplet viscosity, thereby suppressing Marangoni convection. Figure 6(b) shows the variation of surface tension and viscosity ratios with concentration expressed as a fraction of CMC. Both surface tension and viscosity values are non-dimensionalized by considering their initial values at the 0 CMC surfactant concentration. It can be clearly seen that the addition of surfactant dramatically reduces the surface tension, with the surface tension eventually becoming constant beyond 1 CMC. The near-constant value of surface tension indicates the formation of micelles in solution. Micelle formation is highly likely because the working temperature is above the SDS Krafft temperature[43,44], given that the CMC is ~ 8 mM. The formation of micelles increases the viscosity of the aqueous solution. It can also be observed that there is a sharp rise in viscosity from 0.5 CMC. Moreover, further increase in surfactant concentration leads to the continuous rise in viscosity. The increase in viscosity impedes advective Marangoni flow. Therefore, higher surfactant concentrations are not useful in increasing evaporation due to both viscous dissipation and surfactant crowding at the liquid-vapour interface. It should also be noted that evaporation leads to the loss of water, which in turn increases the surfactant concentration relative to the start of evaporation process. Therefore, the average velocity for the 0.5 CMC case decreases by 54.3% at t* ~ 0.5 compared to t* ~ 0.25. On the contrary, 0.25 CMC concentration case shows a marginal increase in the velocity by 16.7% at t*~0.5 compared to t*~ 0.25. In conclusion, we observed in this subsection that lower surfactant concentrations are effective in improving evaporation rates, whereas higher concentrations suppress advection due to viscous damping. However, the exact nature of advection has not been brought out so far. In the upcoming section, we will decipher the main transport mechanisms during the evaporation of surface-active droplets.

## C. Role of soluto-thermal Marangoni advection

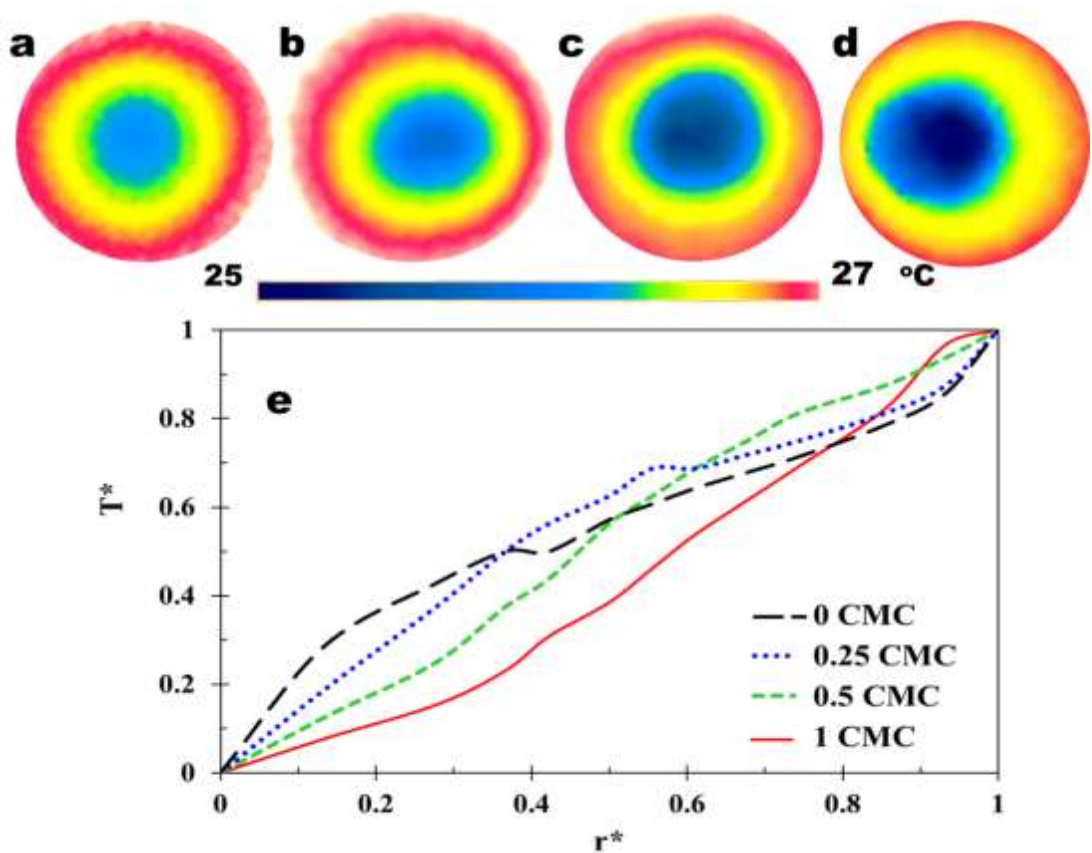

**Figure 6:** Infrared thermography images of the evaporating droplets on a hydrophilic surface for: a) 0 CMC, b) 0.25 CMC, c) 0.5 CMC, and d) 1 CMC concentrations. All the thermal contours here are top-view contours of the droplets. Here, the outer red-coloured regions in the contours represent the substrate. The extracted non-dimensionalized temperature has been shown against the non-dimensional droplet radius in part e (see Figure A4 in appendix for the thermal contours on SHS).

We first analyse the infrared images of evaporating droplets to understand the role of thermal advection. Figure 6 (a-d) shows the variation of temperature over the droplet at various surfactant concentrations for a hydrophilic surface. Infrared thermography is conducted over the same duration as the PIV data shown in Figure 4 to align the time frames for comprehensive analysis. It should be noted that thermal contours shown are top-view contours of droplets. Figure 6(e) shows the non-dimensional variation of temperature ($T^*$) with non-dimensional radius $r^* = r/r_i$, where $r_i$ is the initial radius. Temperature is non-dimensionalized considering maximum temperature ($T_{max}$) and minimum temperature ($T_{min}$) as $T^* = \frac{T-T_{max}}{T_{max}-T_{min}}$. The temperature distribution is calculated from infrared images. It can be observed that the temperature is the lowest around the centre of the droplet. The temperature increases radially outward towards the droplet's periphery in all cases. Water vapour evaporates by extracting energy from the droplet, thereby cooling the droplet. Therefore, a higher evaporation rate should result in an increase in cooler temperatures at the droplet's liquid-vapour interface. Specifically, it can be observed that the temperature remains colder for surfactant-laden droplets, especially for $r^* < 0.25$. It might be attributed to Marangoni advection generated within the droplet, which leads to a higher evaporation rate. However, the droplet temperature is similar for water and surfactant-laden droplets away from the droplet's centre, towards the "rim".

The infrared images for superhydrophobic surface (SHS) are shown in Figure A4 (a, b, c, d) with variation in surfactant concentration. The outer red-coloured regions in the contours represent the substrate. It can be observed that temperatures are overwhelmingly colder with increase in surfactant concentration compared to pure water. Figure A4(e) shows the variation of non-dimensionalized temperature ($T^*$) variation with non-dimensional radius ($r^*$) for SHS substrate, demonstrating lower temperatures at all surfactant concentrations compared to pure water. The temperature decreases as surface concentration increases from 0 to 1 CMC, especially for $0.4 < r^* < 0.8$. It has been observed earlier that the evaporation rate increases (Figure 2c) with an increase in surfactant concentration (CMC) for SHS. Therefore, observed lower temperatures complement the higher evaporation rates.

We need to understand the dominant transport mechanism for the droplet evaporation. Therefore, dimensional analysis is performed to compute the dimensionless numbers. The scaling analysis is conducted to quantify the thermal-solutal advection. Let's consider a sessile drop with instantaneous contact angle ($\theta$), height ($h$), and contact diameter ($d_c$). The instantaneous drop volume can be expressed using spherical cap approximation,

$$V(t) = \frac{\pi h}{6}\left[\frac{3}{4}d_c^2 + h^2\right] \tag{1}$$

We first consider the energy balance for thermal mode driven transport,

Rate of energy change of droplet = Rate of energy change due to thermal conduction across droplet + Rate of energy change due to thermal convection within droplet ± Rate of energy change due to surface interactions of droplet.

The energy changes due to thermal conduction ($\dot{E}_{conduction}$) is written considering $k$, $A_c$, and $\Delta T$ as the thermal conductivity of aqueous solution, base area of the droplet, and temperature difference caused due to the cooling by evaporation, respectively:

$$\dot{E}_{conduction} = k\, A_C\, \frac{\Delta T}{h} \tag{2}$$

Now, the temperature difference across the droplet is expressed considering $\mu$, $h_{fg}$, $\sigma_T = \frac{d\sigma}{dT}$, $C_p$, and $\dot{x}$ as the dynamic viscosity, enthalpy of evaporation, thermal surface tension gradient, heat capacity, and time derivative of the characteristic length, respectively.

$$\Delta T = \sqrt{\frac{\mu h_{fg}}{\sigma_T C_p}(\dot{x})} \tag{3}$$

The droplet pinning for hydrophilic substrate (HP) dominates the evaporation period. Therefore, we take the characteristic length for HP case to be $d_c$ = constant, while $h$ reduces with time. Similarly, droplets on a superhydrophobic substrate (SHS) maintain almost spherical shape. We consider $h$ to be the characteristic length for the SHS. Here, $(\dot{x}) = (\dot{h})$ for HP, and $(\dot{x}) = (\dot{d}_c)$ for SHS. The energy changes due to thermal advection ($\dot{E}_{advection}$) is written considering $\rho$, $U_m$, and $A_s$ as the density of the aqueous solution, spatio-temporally averaged velocity, and surface area of the droplet, respectively:

$$\dot{E}_{advection} = \rho\, C_p U_m \Delta T A_s \tag{4}$$

Here, the surface area of spherical cap is given by $A_s = \pi\left(\frac{d_c^2}{4} + h^2\right) = (A_c + \pi h^2)$

Thus, Equation (4) becomes,

$$\dot{E}_{advection} = \rho\, C_p U_m \Delta T (A_c + \pi h^2) \tag{5}$$

We now take a sessile droplet contact angle with surface tensions of solid-liquid ($\sigma_{SL}$), solid-gas ($\sigma_{SG}$). For the change in energy due to interfacial interactions, as $A_c$ reduces, $A_c\sigma_{SL}$ component reduces, and $A_c\sigma_{SG}$ component increases. Thereby, the net energy change ($E_{surface}$) is:

$$E_{surface} = A_c(\sigma_{SG} - \sigma_{SL}) = A_c\sigma_{LG}\cos\theta = A_c\sigma\cos\theta \tag{6}$$

Now the rate of change of surface energy ($\dot{E}_{surface}$) is thereby given by,

$$\dot{E}_{surface} = \frac{d}{dt}(A_c\sigma\cos\theta) = \frac{d}{dt}\left(\frac{\pi d_c^2}{4}\sigma\cos\theta\right) \tag{7}$$

For hydrophilic case (HP), $d_c$ = constant, hence, $\dot{E}_{surface} = -\frac{\pi d_c^2}{4}\sigma(\sin\theta)\dot{\theta}$

$$\dot{E}_{surface} = -A_c\sigma(\sin\theta)\dot{\theta} \tag{8}$$

From energy balance for the thermally driven transport processes, we can write the Eq. (9) considering the loss of water $\dot{m}$ due to evaporation,

$$\dot{m}h_{fg} = k\, A_C\, \frac{\Delta T}{h} + \rho\, C_p U_m \Delta T(A_c + \pi h^2) - A_c\sigma(\sin\theta)\dot{\theta} \tag{9}$$

We further modify the above equation to get Eq. (10) as

$$\frac{\dot{m}h_{fg}}{k\, A_C\, \frac{\Delta T}{h}} = 1 + \frac{\rho\, C_p U_m \Delta T(A_c + \pi h^2)}{k\, A_C\, \frac{\Delta T}{h}} - \frac{A_c\sigma(\sin\theta)\dot{\theta}}{k\, A_C\, \frac{\Delta T}{h}} \tag{10}$$

Evaluating the 2$^{nd}$ term on RHS as

$$\frac{\dot{E}_{advection}}{\dot{E}_{conduction}} = \frac{\rho\, C_p U_m(A_c + \pi h^2)}{\frac{k\, A_C}{h}} \tag{11}$$

Now, thermal Ma velocity is $U_m = \frac{\sigma_T \Delta T}{\mu}$. Hence, Equation (11) becomes

$$\frac{\dot{E}_{advection}}{\dot{E}_{conduction}} = \left(\frac{\rho\, C_p}{k}\right)\left(\frac{\sigma_T \Delta T}{\mu}h\right)\left(\frac{A_c + \pi h^2}{A_c}\right) \tag{12}$$

Now, considering $Ma_T = \frac{\sigma_T \Delta T}{\alpha\mu}h$ and $\alpha$ to be thermal diffusivity, the Equation (13) can be modified as

$$\frac{\dot{E}_{advection}}{\dot{E}_{conduction}} = (Ma_T)\left(1 + \frac{\pi h^2}{A_c}\right) \tag{13}$$

Evaluating the 3rd term on RHS as

$$\frac{\dot{E}_{surface}}{\dot{E}_{conduction}} = \frac{A_c \sigma (\sin\theta)\dot{\theta}}{k A_c \frac{\Delta T}{h}} = \frac{\sigma (\sin\theta)\dot{\theta}}{k \Delta T} h \tag{14}$$

Now, using the Capillary number $(Ca) = \frac{\mu V}{\sigma} = \frac{\mu(\dot{\theta}h)}{\sigma}$, we get $\dot{\theta}h = \frac{\sigma Ca}{\mu}$. Also, use $\sigma = \sigma_T \Delta T$. So, now Equation (14) can be written as

$$\frac{\dot{E}_{surface}}{\dot{E}_{conduction}} = \left(\frac{\sigma_T \Delta T}{\mu \alpha} h\right)(\sin\theta)\left(\frac{\alpha \sigma Ca}{k h \Delta T}\right) \tag{15}$$

Equation (15) can be modified by using Marangoni number ($Ma_T$) as

$$\frac{\dot{E}_{surface}}{\dot{E}_{conduction}} = (Ma_T)(\sin\theta) Ca \left(\frac{h_{fg}}{C_p \Delta T}\right)\left(\frac{\frac{\sigma}{h}}{\rho h_{fg}}\right) \tag{16}$$

Now, using $\left(\frac{h_{fg}}{C_p \Delta T}\right)$ = latent heat of vaporization/ sensible heat = 1/evaporative Jakob number ($Ja_e$), and $\left(\frac{\frac{\sigma}{h}}{\rho h_{fg}}\right)$ = surface energy change/ latent thermal energy = Evaporation number ($Evp$)

$$\frac{\dot{E}_{surface}}{\dot{E}_{conduction}} = (Ma_T)(\sin\theta) Ca \left(\frac{1}{Ja_e}\right)(Evp) \tag{17}$$

Hence, the equation (10) is modified in dimensionless form for hydrophilic surface to get Eq. (18) by utilizing Eq. (17) and Eq. (13). Here, thermal Marangoni number evaluates the strength of thermal Marangoni led advection relative to the diffusive transport, for aqueous surface-active droplets. Capillary number quantifies the dominance of viscous forces relative to surface tension force, while Evaporation number compares the energy scales for surface and latent heat of evaporation.

$$\frac{\dot{m}h_{fg}}{k A_C \frac{\Delta T}{h}} = 1 + (Ma_T)\left(1 + \frac{\pi h^2}{A_c}\right) - (Ma_T)Ca\left(\frac{1}{Ja_e}\right)(Evp)(\sin\theta) \tag{18}$$

It must be noted that $Ma_T$ expression is same for SHS surface, and capillary number $Ca = \frac{\mu V}{\sigma} = \frac{\mu(\dot{d}_c)}{\sigma}$ and $Evp = \left(\frac{\frac{\sigma}{d_c}}{\rho h_{fg}}\right)$.

Droplet evaporation creates a cooler liquid-vapour interface relative to the aqueous solution adjacent to the substrate, thereby inducing thermal gradients within the droplet. These thermal gradients can trigger buoyancy-driven Rayleigh-Bénard convection. Therefore, it is important to do scaling analysis and quantify the model parameters to compare Rayleigh convection to Marangoni thermocapillary convection. Rayleigh number is written considering $\beta$, $\Delta T_{Ra}$, and $g$ as the thermal expansion coefficient, temperature difference triggering convection, and acceleration due to gravity, respectively as

$$\Delta T_{Ra} = \left( \frac{\mu h_{fg} \dot{\theta}}{\rho g \beta d_c C_p} \left( 1 + \left( \frac{h}{\frac{d_c}{2}} \right)^2 \right) \right)^{\frac{1}{2}} \tag{19}$$

$$u = \frac{\rho g \beta \Delta T_{Ra} d_c^2}{4\mu} \tag{20}$$

$$Ra = \frac{d_c^2}{8\alpha} \left( \frac{\rho g \beta d_c h_{fg} \dot{\theta}}{\mu C_p} \left( 1 + \left( \frac{h}{\frac{d_c}{2}} \right)^2 \right) \right)^{\frac{1}{2}} \tag{21}$$

It needs to be determined what the dominant mode of convection is among thermal Marangoni and Rayleigh convection. Now, we use the stability analysis from Nield[45] and Davis[46] for expressing the advection stability

$$\frac{Ma}{Ma_s} + \frac{Ra}{Ra_c} = 1 \tag{22}$$

where $Ra_c$ and $Ma_s$ are the critical Rayleigh and solutal Marangoni numbers. The critical Marangoni number ($Ma_s$) can be taken ~ 80[45,47] and critical Rayleigh number ($Ra_c$) ~ 1708 from Chandrasekhar' analysis. The stability analysis involving variation of thermal Marangoni number with Rayleigh number, identifies the dominant mode of advection. Figure 7(a) shows the variation of thermal Marangoni number with Rayleigh number, considering both variations in concentration and substrate wettability. Two important regimes are drawn by joining $Ra_c$ ~1708 – $Ma_s$ ~ 80 (Nield[45]) and $Ma_s$ ~ 52 (Davis[46]). It is important to consider that values lying above the Nield line portray stable advection, whereas values lying below the Davis line depict unstable advection. Moreover, the values lying between Davis and Nield lines depict partially stable advection.

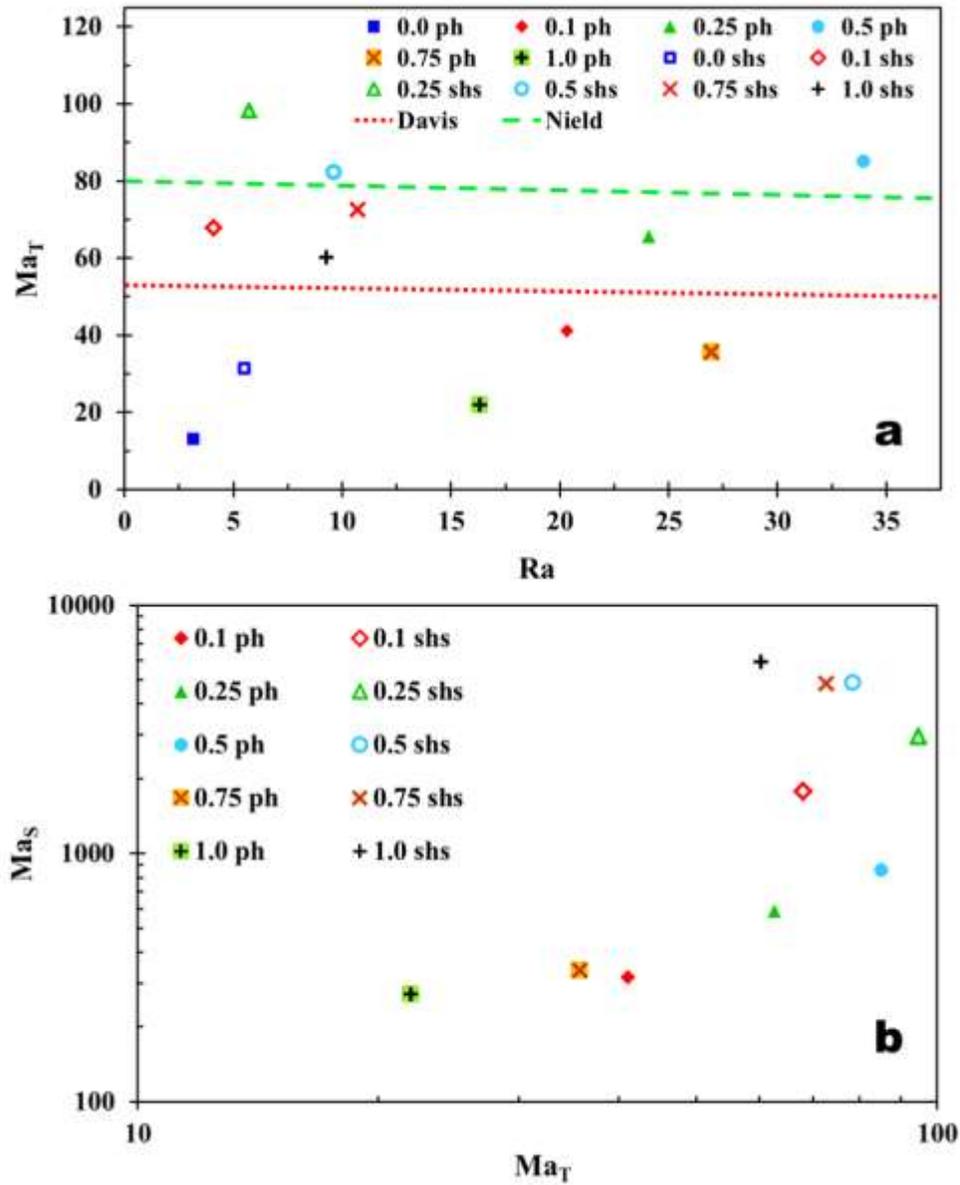

**Figure 7.** (a) Variation of thermal Marangoni number ($Ma_T$) against Rayleigh number ($Ra$) for both hydrophilic (ph) and superhydrophbic surfaces (shs) with variation in concentration; (b) Variation of solutal marangoni number ($Ma_s$) against thermal Marangoni number ($Ma_T$).

It can be seen that the Rayleigh number reaches a maximum of ~ 35, which is significantly less than the critical Rayleigh number. Therefore, advection due to buoyancy-driven Rayleigh flow can be neglected. For the hydrophilic surface, the thermal Marangoni number increases with an increase in SDS concentration from 0 to 0.1 CMC, but advection remains unstable. Advection remains partially or intermittently stable at 0.25 CMC and becomes fully stable at 0.5 CMC. However, further increase in the surfactant concentrations leads to the decrease in Marangoni number and rise in instability. The strong thermal Marangoni advection at 0.5 CMC directly complements the highest evaporation rates observed experimentally. Moreover, the rise and fall of Marangoni thermal advection complement the evaporation rates observed. Further, it can be clearly seen that the Marangoni number continues to rise with increasing surfactant concentration for the superhydrophobic case up to 0.25 CMC, highlighting the dominance of thermal Marangoni convection. The Marangoni thermal advection is found stable only at 0.25 CMC and 0.5 CMC for SHS, whereas the stability only occurs at 0.5 CMC for the hydrophilic substrate. Therefore, Marangoni thermal advection is stable only at a very few concentrations and insufficient to explain the trends observed during the evaporation. Therefore, we need to explore further Marangoni solutal advection to understand the dominant transport mechanism to explain the evaporation rates.

Next, we shall focus on energy balance arising from mass transfer processes to develop a scaling model. We write the energy ($\dot{E}_{diffusion}$) by considering diffusive mass $\dot{m}_{diffusion}$

$$\dot{E}_{diffusion} = \dot{m}_{diffusion} C_p \Delta T = D A_c \frac{\Delta C}{h} C_p \Delta T \tag{23}$$

Where, $\Delta C$ is the concentration difference between the interface and bulk. $\Delta C$ is further written similarly to $\Delta T_{Ma_T}$ as $\Delta C = \sqrt{\frac{\rho\mu}{\sigma_c}(\dot{x})}$. Here, $(\dot{x}) = (\dot{h})$ is used for hydrophilic, and $(\dot{x}) = (\dot{d}_c)$ for SHS. The species advection ($\dot{E}_{advection}$) can be written considering solutal Marangoni velocity ($U_C$) as $\dot{E}_{advection} = \dot{m}_{advection} C_p \Delta T = (U_C \Delta C A_s) C_p \Delta T$. So, $\dot{E}_{advection}$ is given by Equation (24)

$$\dot{E}_{advection} = (U_C \Delta C)(A_c + \pi h^2) C_p \Delta T \tag{24}$$

We write the $\dot{E}_{surface} = -A_c \sigma (\sin\theta)\dot{\theta}$ using the similar calculation procedure used in Eq. (8). Now, energy balance for solutal driven flow is given by Equation (25).

$$\dot{m} C_p \Delta T = D A_c \frac{\Delta C}{h} C_p \Delta T + (U_C \Delta C)(A_c + \pi h^2) C_p \Delta T - A_c \sigma (\sin\theta)\dot{\theta}$$

$$\frac{\dot{m}}{D\,A_C\,\frac{\Delta C}{h}} = 1 + \frac{(U_C\Delta C)(A_c + \pi h^2)}{D\,A_C\,\frac{\Delta C}{h}} - \frac{A_c\sigma(\sin\theta)\dot{\theta}}{D\,A_C\,\frac{\Delta C}{h}\cdot C_p\Delta T} \tag{25, 26}$$

We utilize solutal *Ma* velocity as $U_C = \frac{\sigma_C \Delta C}{\mu}$ and $Ma_s = \frac{\sigma_C \Delta C}{\mu D} h$ to obtain the second term on RHS as

$$\frac{\dot{E}_{advection}}{\dot{E}_{diffusion}} = \frac{U_S(A_c + \pi h^2)}{D\,A_C/h} = (Ma_s)\left(1 + \frac{\pi h^2}{A_c}\right) \tag{27}$$

Now, using $Ca = \frac{\mu V}{\sigma} = \frac{\mu(\dot{\theta}h)}{\sigma}$ and $\sigma = \sigma_C \Delta C$ to evaluate the 3$^{rd}$ term on RHS

$$\frac{\dot{E}_{surface}}{\dot{E}_{diffusion}} = \frac{A_c\sigma(\sin\theta)\dot{\theta}}{D\,A_C\,\frac{\Delta C}{h}\cdot C_p\Delta T} = \frac{\sigma(\sin\theta)\dot{\theta}}{D\,\Delta C\cdot C_p\Delta T}h = \left(\frac{\sigma}{\Delta C\cdot C_p\Delta T\cdot h}\right)(\sin\theta)\left(\frac{\sigma_c\Delta C}{\mu D}h\right)\cdot Ca \tag{28}$$

Eq. (28) is further modified using $\left(\frac{h_{fg}}{C_p\Delta T}\right)$ = latent heat of vaporization/ sensible heat = 1/evaporative Jakob number ($Ja_e$), $\left(\frac{\sigma/h}{\rho h_{fg}}\right)$ = surface energy change/ latent thermal energy = Evaporation number (*Evp*), and $\left(\frac{\rho}{\Delta C}\right) = \left(\frac{\text{mass of solution}/\text{Volume}}{\text{mass of solute}/\text{Vol}}\right)$ = non-dimensional mass ratio of solution to solute = 1/dilution ratio (δ)

$$\frac{\dot{E}_{surface}}{\dot{E}_{diffusion}} = (Ma_s)\left(\frac{1}{\delta}\right)(\sin\theta)\,Ca\left(\frac{1}{Ja_e}\right)(Evp) \tag{29}$$

Hence, Eq. (26) is modified using Eq. (27) and Eq. (29) for hydrophilic surface as

$$\frac{\dot{m}}{D\,A_C\,\frac{\Delta C}{h}} = 1 + (Ma_s)\left(1 + \frac{\pi h^2}{A_c}\right) - (Ma_s)\left(\frac{1}{\delta}\right)Ca\left(\frac{1}{Ja_e}\right)(Evp)(\sin\theta) \tag{30}$$

*Ma$_s$* expression is similar to the hydrophilic case, whereas $Ca = \frac{\mu V}{\sigma} = \frac{\mu(d_c)}{\sigma}$ and $Evp = \left(\frac{\frac{\sigma}{d_c}}{\rho h_{fg}}\right)$ for SHS. Equation (30) shows that the solutal Marangoni number is important for evaluating mass transfer-related advection in the evaporating droplet. SDS surfactant molecules have a natural tendency to accumulate at the interface due to their amphiphilic nature. It is highly likely that SDS molecules have different concentrations at the interfaces compared to the bulk droplet. Adsorption-desorption of surfactant molecules during the evaporation process disturbs the localized surface tension gradients.

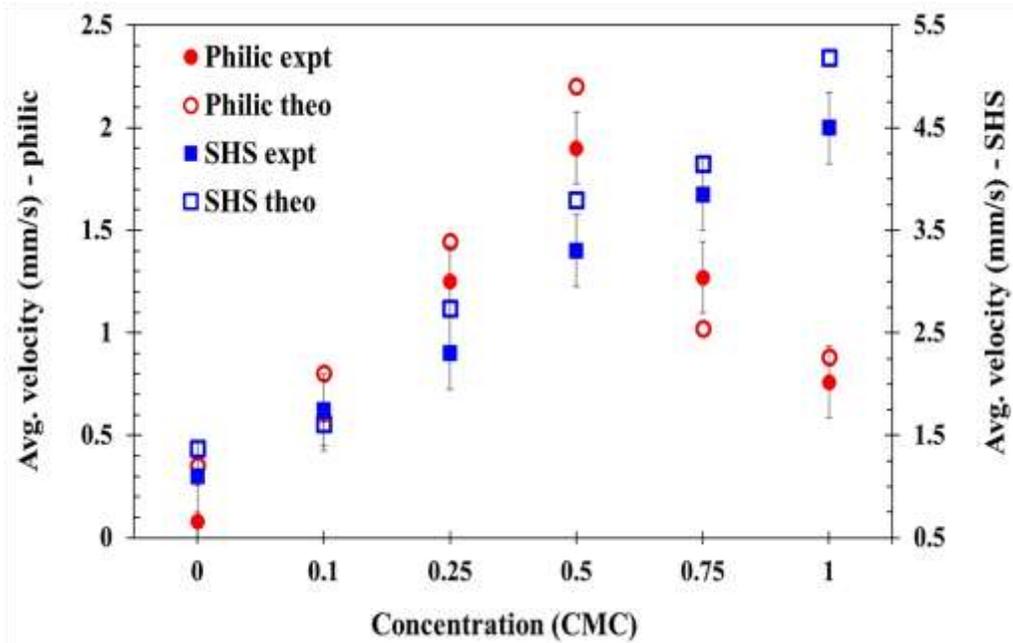

**Figure 8.** Comparison between experimentally obtained average velocities (expt) and predicted Marangoni solutal velocities (theo) for superhydrophobic (SHS) and hydrophilic surfaces (Philic). The comparisons between velocities is demonstrated with variation in SDS concentration as a fraction of critical micelle concentration (CMC).

The evaporation process leads to a change in the transient surfactant concentration in the droplet due to loss of water content. However, droplet evaporation rates are non-uniform over the droplet. Therefore, surfactant concentration increases non-uniformly in the droplet and leads to the surface tension gradients. It is expected that surfactant molecules will have a higher concentration at the outer periphery due to the capillary flow for the pinned droplet. A radially outward higher surfactant concentration would lower the surface tension locally. Therefore, the surface tension would be higher at the apex and lower at the outer periphery near the substrate, leading to solutal Marangoni advection and causing surface fluid flow radially towards the droplet's centre. The combination of inward Marangoni flow and outward evaporation-driven flow can form a vortex or circulation loop[48,49]. Figure 7(b) shows the variation of solutal Marangoni number versus thermal Marangoni number. $Ma_s$ proliferates with an increase in surfactant concentration for the hydrophilic substrate. For instance, $Ma_s$ reaches ~1000 at 0.5

CMC, whereas $Ma_T$ is only ~ 90. Therefore, the dominant mechanism is Marangoni solutal advection, responsible for fluid transport within the droplet. However, the $Ma_s$ decreases at 0.75 and 1 CMC compared to the 0.5 CMC. The decrease can be attributed to surfactant crowding and the rise in viscous resistance. The solutal advection for the superhydrophobic surface increases rapidly with increase in surfactant concentration. Interestingly, $Ma_s$ reaches close to 7000 at 1 CMC while $Ma_T$ remains only ~ 60. Therefore, solutal advection is responsible for enhanced evaporation rates and internal advection. We can also conclude that $Ma_T$ has a minor role in advection, and buoyancy-driven Rayleigh flow can be neglected. The SHS surface provides a higher interfacial area, which prevents surfactant crowding and allows higher solutal advection. Now, we have established that solutal Marangoni advection is the dominant transport mechanism. We will further analyze the flow velocities from both experimental and theoretical perspectives to reinforce the conclusion regarding the dominant transport mechanism.

The experimental spatio-temporally averaged velocities from PIV are plotted to compare with solutal Marangoni velocities obtained from scaling analysis in Figure 8. The velocities are compared across the range of SDS surfactant concentration as a function of CMC. It shows a consistent match between the predicted and experimental values. Both theoretical and experimental velocities increase up to 0.5 CMC for the hydrophilic surface due to the enhancement in solutal advection generated by surfactants. The velocities decrease at 0.75 CMC and 1 CMC, possibly due to surfactant crowding at the interface and viscous damping against advection. The experimental and theoretical velocities increase with an increase in surfactant concentration for the SHS substrate. The analysis of variation in flow behaviour with substrate characteristics has already been covered in earlier discussions. Overall, Figure 8 reinforces that solutal Marangoni advection is the dominant transport mechanism in aqueous surfactant droplets.

### D. Deposition Patterns

Figure 9 shows the captured microscopic images to demonstrate the deposition pattern for (a) CTAB and (b) SDS surfactants. The images are shown at surfactant concentrations of 0.25 CMC, 0.5 CMC, and 1 CMC. Specifically, the images show formation of a distinct band of deposited surfactant particles towards the edge of droplet or a "rim-type" deposition. The deposition away from the "rim or ring" type deposition appears to be scanty. The "rim-structures" formation images are also shown in the appendix in Figure A5 at 50 μm and 200 μm scales for CTAB, and SDS surfactants, respectively for better visualization. We have also quantified and plotted the variation of rim's width with variation in surfactant concentration (Figure 10). It can be observed that the "rim" feature is very dominant, especially for SDS-laden droplets.

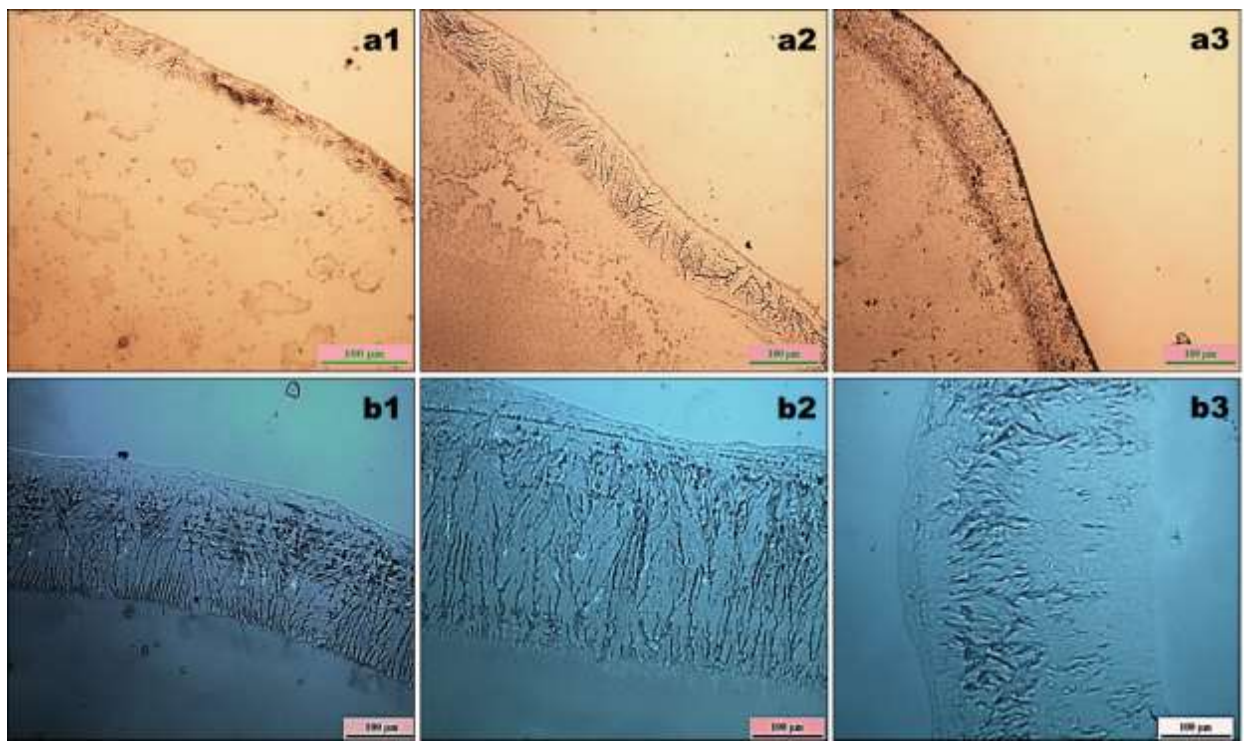

**Figure 9.** Formation of rims after drying of the droplets for (a) Cetyltrimethylammonium bromide (CTAB) aqueous solution at a1) 0.25 CMC, a2) 0.5 CMC, a3) CMC, and (b) Sodium Dodecyl Sulphate (SDS) aqueous solution at b1) 0.25 CMC, b2) 0.5 CMC, b3) 1 CMC (100 µm scale).

The extent of formation of rim is affected by the outward radial flow and inward solutal Marangoni flow. For a pinned evaporating droplet in CCR mode, the fluid flows radially to replenish the loss of water due to evaporation. This radial outward flow deposits particles at the droplet's periphery, known as the coffee ring effect[41]. However, surfactant molecules can generate counter Marangoni flow to suppress or eliminate the effect. It must be remembered that fluid movement not only can carry suspended particles but also carries along the surfactant molecules. The higher surfactant concentration at the edges might also promote surfactant adsorption at the interface, which not only locally reduces the interfacial tension but also affects droplet pinning. The high surface tension at the droplet apex pulls fluid inwards, away from the three-phase contact line. Still et al.[48] demonstrated that polystyrene particles could not reach the edge of the droplet because they got trapped in the Marangoni eddy generated near the edge of the droplet.

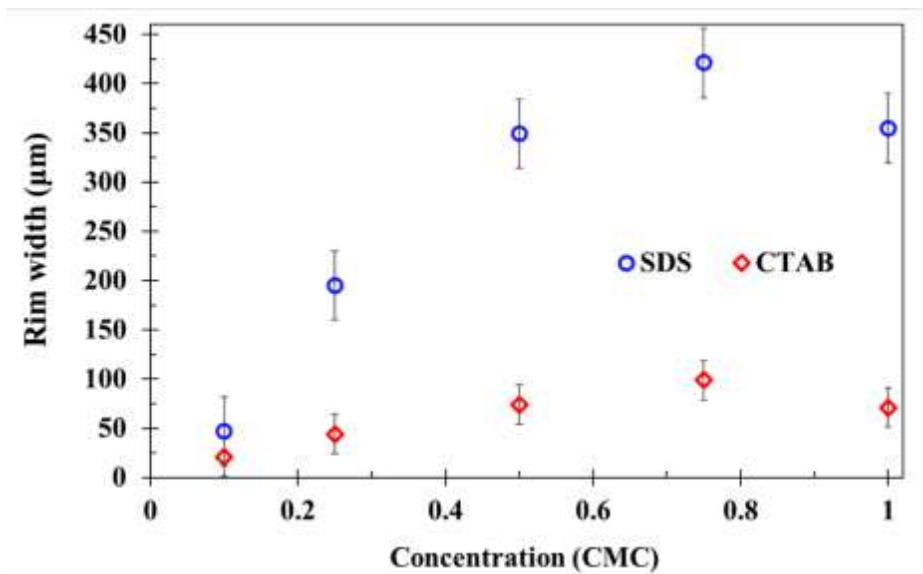

**Figure 10.** Variation of rim width for SDS and CTAB surfactants with the change in concentration.

Furthermore, the authors found that resistance to the radially outward capillary flow by the Marangoni eddy led to the formation of a distinct rim, separated from the rest of the droplet. Therefore, the large width of the deposited rims might be due to the partial suppression of the coffee ring effect. The "Marangoni eddy" creates resistance to capillary outward flow and diminishes the surge in surfactant-molecule concentration at droplet edges. Surface tension gradients drive fluid flow towards the centre, which may depict "levelling up" or make particle deposition relatively more uniform. For instance, Kajiya et al.[50] found that the addition of 5 wt% F489 surfactant to polymer solution droplets made deposition height more uniform spatially. It can also be noted that rim widths for SDS aqueous solution are much higher than those for CTAB. The rim width for SDS and CTAB reaches the maximum value ~ 425 µm, and 100 µm, respectively (Figure 10). The higher rim width of SDS is the combination of two factors. Firstly, the concentration of SDS surfactant molecules at CMC is higher than that of

CTAB[51]. Therefore, there are higher molecules available for deposition at the rim. Secondly, SDS droplets evaporate more rapidly than CTAB droplets, implying that advection within the CTAB droplet is weak. The weaker advection might decrease the width of the rim, and make surfactant deposition highly concentrated at the edges due to a decline in the "levelling effect". The effect of a decrease in advective velocities is also visible in the SDS solution deposited pattern. It can be argued that an increase in surfactant concentration should monotonically increase the rim's width due to the availability of more surfactant molecules. However, the rim's width at 1 CMC is lower than at 0.75 CMC. The decrease in rim's width at 1 CMC is surprising because there are higher surfactant molecules available for deposition. Therefore, the increase in concentration at 1 CMC is not able to compensate for the decreased advection at 1 CMC, as average advective velocities decrease after 0.5 CMC. Therefore, the strength of Marangoni solutal advection plays a crucial role in the final deposition pattern and the rim's width.

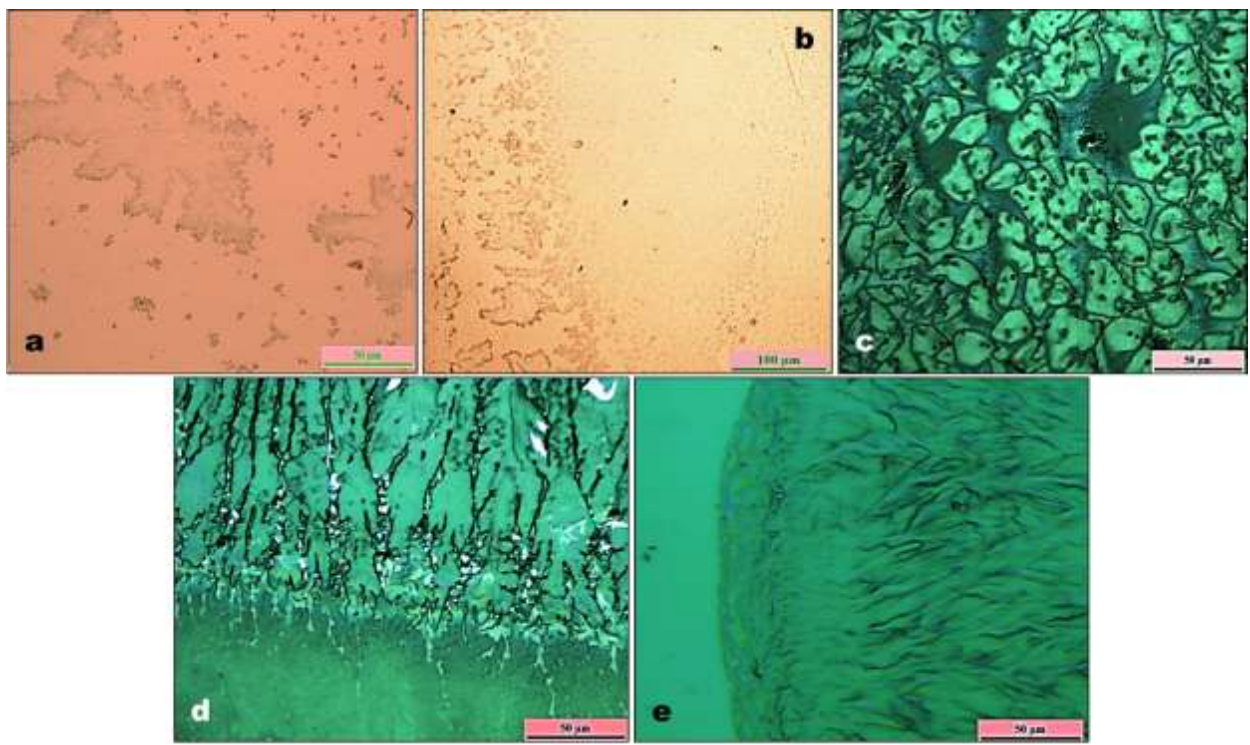

Figure 11: Morphology of the deposition pattern of the aqueous SDS solution droplet. (a,b) Patch-type depositions away from rim structure, (c) Cell-type depositions, (d,e) Fingering patterns within the deposited rim structure

Figure 9 also shows deposited rim has embedded fingers or a waviness pattern after drying. The fingering pattern within the rim structure might be due to visco-capillary instability, arising from the competition between fluid driven by surface-tension gradients and viscous resistance. Surface tension drives the motion, but viscous effects spread the fluid slowly, allowing perturbations to grow. However, it is difficult to pinpoint the exact nature of instability at this stage. It might also be possible that Marangoni-led flow also creates instabilities. The presence of instabilities has been reported during the drying of the surfactant-laden droplets. Dier et al.[52] observed instabilities during the drying of droplet containing non-ionic surfactant Triton X-100. The authors found that instability has a thermocapillary nature, impacted by both thermal and surfactant-led Marangoni effects. Tiwari et al.[53] observed the drying pattern of the droplets containing surfactant molecules and PS particles, and found the formation of "fingering" patterns. The authors reasoned that particles moving due to Marangoni eddy experienced non-balanced "capillary meniscus forces". These imbalanced forces led to the clustering of the particles, which eventually deposited to form the fingering pattern. Figure 11 shows the morphology of the deposits from the drying of SDS droplets, affected by instabilities. The deposition away from the rim are shown in Figure 11 (a, b, c) consisting of the patch or cell-type deposits. The multiple cellular-type deposits in Figure 11(c) might indicate organised fluid structures prior to deposition. Figure 11(d-e) demonstrates the fingerings or waviness within the deposited rims at 50 µm scale. It must be also emphasized that deposition pattern is also affected by "rush hour effect". Marín et al.[54] explained the "rush hour effect" as a sudden increase in velocity or a "temporal singularity in velocity" in the last moments of drying. They found that the deposition pattern was ordered in the earlier stage due to Brownian motion, and disorder in the deposition increases due to a sudden spike in velocity. Therefore, effect of fluid motion is complex on the drying pattern of the droplet.

SDS solution droplets also demonstrated a very interesting phenomena of "multiple rings" during deposition. Figure 12(A) shows the formation of multiple rings at a1) 0.25 CMC, a2) 0.5 CMC, and a3) CMC. Obviously, the formation of multiple rings shows presence of SDS molecules in aqueous solution affects the depinning of the contact line. Specifically, the multiple rings are due to the "stick and slip" behaviour of the contact line in the droplet. Figure 12(B) perfectly captures stick-slip behaviour by measuring the contact line (CL) velocity. The zero contact line velocity shows the pinning of contact line or "sticking" behavior whereas the sudden spike in contact line velocity shows the depinning of contact line. The contact line at 0 CMC remains pinned and does not exhibit any major spike in CL velocity, confirming the lack of stick-slip behaviour. Now, the increase in concentration to 0.25 CMC leads to multiple spikes in CL velocity between 800 s and 1000 s. These multiple spikes correspond to the formation of a multiple-ring pattern, as indicated by arrows in Figure B. It can also be observed that this "multiple spike behaviour" occurs at all surfactant concentrations. The final dip in CL velocity is due to the complete drying of sessile droplet. Moreover, the "final dip" occurs more quickly at 0.5 CMC than at the rest of the surfactant concentration, indicating a faster evaporation rate. It is important to understand the reasons for the peculiar drying behaviour displayed by the aqueous surfactant droplets. The contact line gets pinned due to factors such as chemical inhomogeneities and surface defects. It is mathematically complex to quantify the relation between surface characteristics and pinning. Now, the droplet pinning phase might lead to surfactant molecules' aggregation near the contact line and subsequent precipitation as droplet evaporation increases. Moreover, it is highly likely that surfactant molecules interact with the substrate and affect adhesive forces. Shao et al.[55] suggested that surfactants could be electrostatically attracted to the bare solid-vapour interface, which decreases the solid-vapour interfacial tension and increases hydrophobation of the interface. The hydrophobation of the hydrophilic surface will assist depinning and contact-line receding. Ultimately, the weakening of pinning forces leads to the contact line movement or depinning. Some of the surfactant molecules remain attached to the substrate, aggregate and deposit as the contact line recedes during the depinning event. The receding contact line moves inward, and further pinning will result in the formation of a new ring. Therefore, the occurrence of cycles of pinning, precipitation, and depinning results in a multiple-ring pattern.

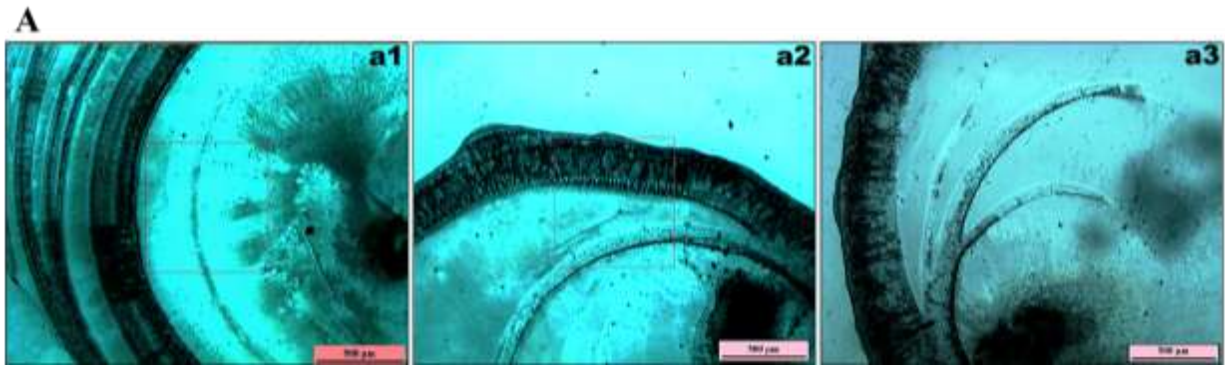

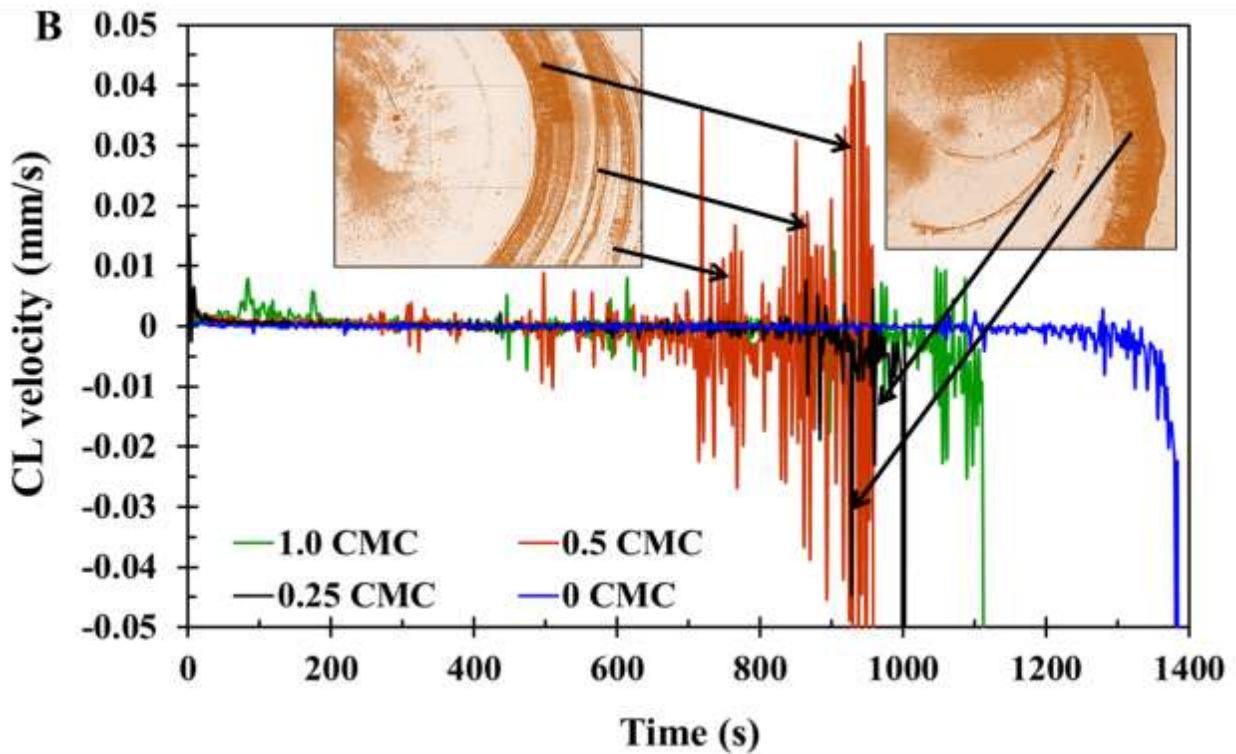

**Figure 12:** (A) Multiple rings formation during deposition of surfactant molecules at a1) 0.25 CMC, a2) 0.5 CMC, a3) CMC; (B) Contact line (CL) velocity of the droplet to demonstrate the multiple ring patterns observed from the images. The arrow marks in the CL velocity plot indicate the spikes due to the formation of the multiple ring pattern.

Popov[42] remarked that pinning force is "relatively insensitive" to the contact angle and depends upon the "materials involved". Therefore, the pinning force depends upon the interaction between the aqueous surfactant molecules and the substrate in the present case. Now, the decrease in contact angle due to evaporation leads to an increase in depinning force, while the pinning force remains relatively constant. The continuous decrease in contact angle crosses the threshold value, allowing the depinning force to overcome the pinning force. Li et al.[56] described the driving force ($F_{df}$) for contact line receding as $F_{df} = \gamma(\cos\theta - \cos\theta^e)$, where $\gamma$, $\theta$, and $\theta^e$ are the interfacial tension of the gas-liquid interface, the dynamic contact angle, and the contact angle at equilibrium, respectively. It must be emphasized that $\gamma$ reduces due to the presence of surfactant molecules, and contact angles ($\theta, \theta^e$) are affected due to surfactant molecules' adhesion to the substrate. Furthermore, the interfacial tension continuously changes during evaporation due to changes in surfactant concentration. The driving force expression is a simplistic representation of the complex phenomena of receding. The contact line would recede if the driving force for receding exceeded the droplet pinning force. It is worth noting that stick-slip behaviour also influences the capillary flow because it is dependent on the evaporation rate, which itself is proportional to the droplet's radius. Therefore, capillary flow's dependence on wetting area makes it sensitive to contact line slipping. The Marangoni fluid flow inside the droplet may also indirectly contribute to the formation of multiple ring patterns by ensuring that most of the surfactant concentration is not concentrated at the droplet edges. This allows sufficient surfactant concentration away from the edges to supersaturate and precipitate. Still et al.[48] also suggested that particle concentration is higher in the Marangoni vortex due to non-attachment to the drop's edge, which ultimately helps create multiple ring structures.

However, it is difficult to generalize and extend the effect of SDS surfactant on the aqueous droplet evaporation to other surfactants without careful examination. Surfactants may exhibit different behaviour depending on their interactions with aqueous solutions and substrates. We observe that multiple ring patterns are insignificant for CTAB solution droplets. We have already shown that both evaporation rate and rim thickness are lower for CTAB. The divergence in the behaviour of SDS and CTAB aqueous solutions during droplet evaporation stems from both their inherent chemical structures and thermophysical properties. In conclusion, the presence of surfactant molecules influences both the advection pattern and the adsorption process. This study uniquely relates advection within the droplet to deposition patterns.

## 4. CONCLUSIONS

The present work studies the evaporation kinetics and deposition pattern of sessile surfactant-laden droplets on hydrophilic surfaces and SHS. The work aims to relate the observed evaporation rates and deposition patterns to the transport phenomena inside the droplet. The major conclusions from the present work are as follows:

- Experimental and theoretical analysis indicate that solutal Marangoni convection is the dominant transport mechanism for enhanced advection inside the droplet by surface-active droplets. This surface tension gradient-driven flow shears the vapour diffusion layer surrounding the droplet, resulting in enhanced evaporation.
- The concentration of surfactant relative to CMC modulates the droplet evaporation lifetimes, decay rates and deposition pattern. For instance, the evaporation rate increased up to 0.5 CMC and declined thereafter for the hydrophilic substrate. It was found that both Marangoni solutal and thermal velocities decrease beyond 0.5 CMC with a simultaneous dramatic increase in dynamic viscosity. Therefore, the effectiveness of surfactant droplets depends upon the competition between the advection and viscous resistance.
- The computed non-dimensionalized geometric parameters provided a subtle but robust interplay between surfactant-induced interfacial dynamics and substrate wettability in governing droplet evaporation. For both hydrophilic and hydrophobic substrates, the normalized volume V* decreases approximately linearly over a substantial fraction of the droplet lifetime, suggesting evaporation dominated by a quasi-steady vapour diffusion process. However, the droplet decay slope rates and lifetimes rely strongly on SDS concentration and surface chemistry. The non-dimensionalized contact diameter and contact angle indicated the dominant mode of evaporation. Constant contact radius (CCR) modes were found to be dominant in hydrophilic substrates, where constant contact angle (CCA) was dominant during evaporation at the SHS surface.
- The substrate wettability impacts the evaporation rates, with the sessile droplet having a higher evaporation rate on the hydrophilic substrate compared to the super hydrophobic substrate (SHS). The evaporation rate is lower for SHS due to reduced solid–liquid contact area and higher thermal resistance related to the air cushion beneath the droplet.
- The aqueous surfactant solution droplets demonstrated the sudden spike/ fluctuations in contact line velocities during the evaporation process, which complements the multiple rings pattern observed from the drying pattern images. The enhanced Marangoni flows assist in the formation of multiple ring patterns or "stick-slip behaviour". However, pinning and depinning of the contact line are strongly influenced by changes in adhesion and cohesive forces arising from surfactant adsorption/desorption.

- The "rim structure" formed due to surfactant molecules deposition is influenced by both fluid transport and surfactant characteristics. Rim thickness is significantly higher for SDS than for CTAB solution droplets from 0.1 CMC to 1 CMC due to a higher surfactant concentration at the CMC and superior advection within the droplet. The deposition pattern displayed the formation of fingers within the deposited rim structure. The fingering pattern is attributed to fluid flow instabilities.

Overall, the article demonstrates that surfactant-mediated interfacial dynamics and substrate wettability jointly modulate the evaporation of sessile droplets, through their impact on contact-line behaviour, internal flow strength, and vapour-transport scaling. The internal droplet thermo-fluidics impacts both the evaporation and deposition patterns. The observations from the current study have important implications for tuning the deposition patterns.

## APPENDIX

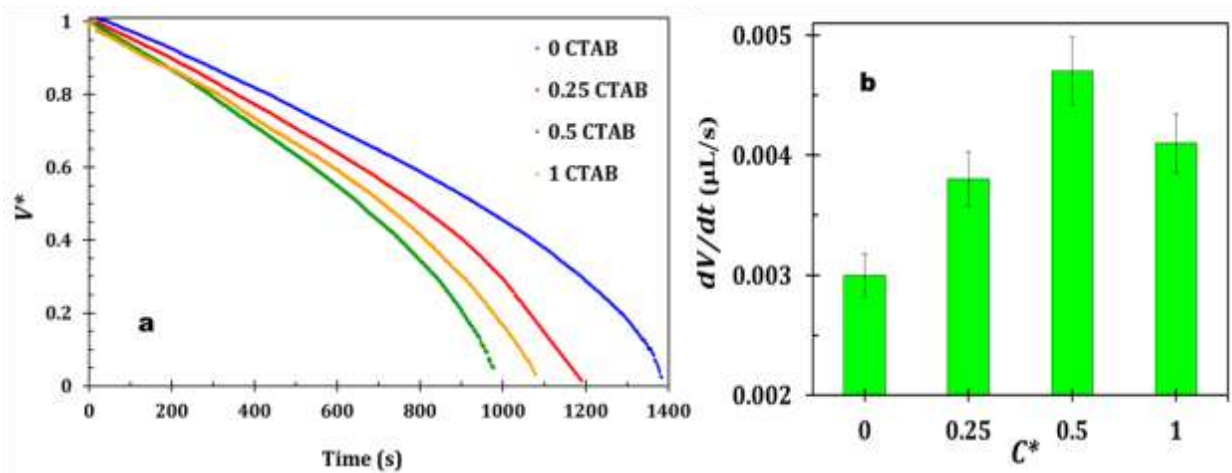

**Figure A1:** a) Non-dimensional droplet volume (V*) vs. time plots for CTAB solution droplets of different concentrations (expressed as fraction of the CMC) over hydrophilic surface, and b) corresponding temporal rate of droplet volume decay (for the linear regimes in part a) as function of CTAB concentration.

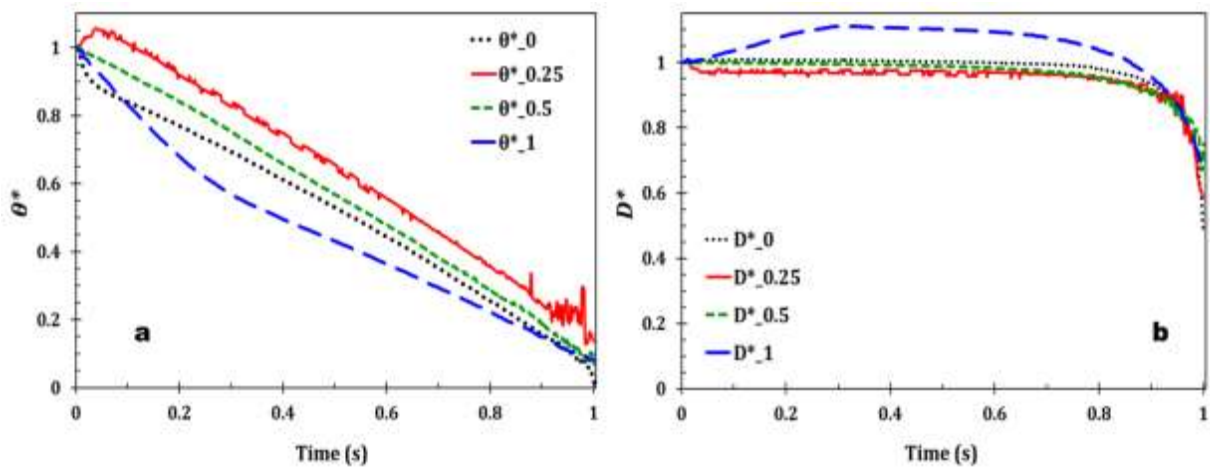

**Figure A2:** (a) Non-dimensional contact angle with time plots and (b) the corresponding non-dimensional wetting diameter with time plots for evaporating sessile droplets of different CTAB initial concentrations.

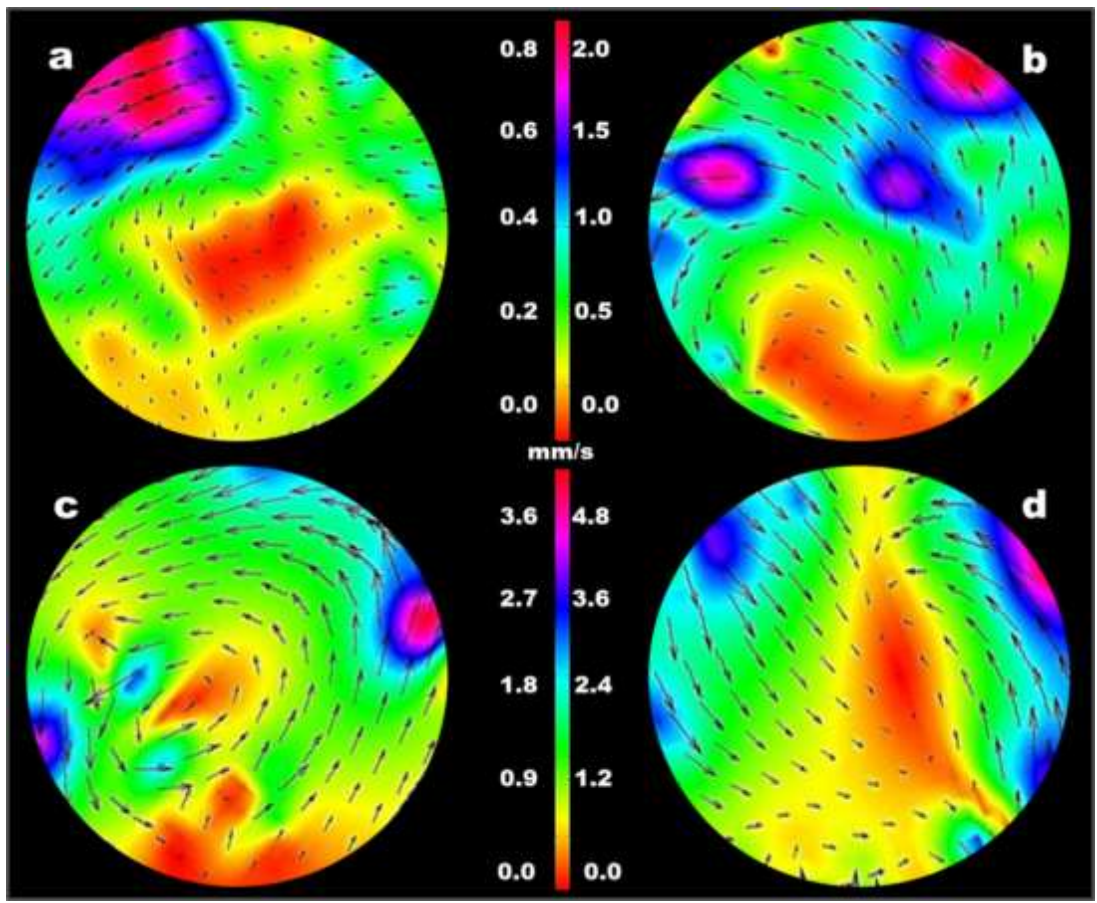

**Figure A3:** Temporally averaged velocity contours (from analyses of PIV experiments data) for SDS solution droplets on SHS: a) 0 CMC, b) 0.25 CMC, c) 0.5 CMC, and d) 1 CMC concentrations. All the contours here are side-view contours of the droplets. The regions of the droplet near the liquid-gas and liquid-solid interfaces are not shown, and the major central portion of the droplet has been used for the PIV analysis (to avoid the stray velocity vectors which are common near interfaces of such microfluidic systems).

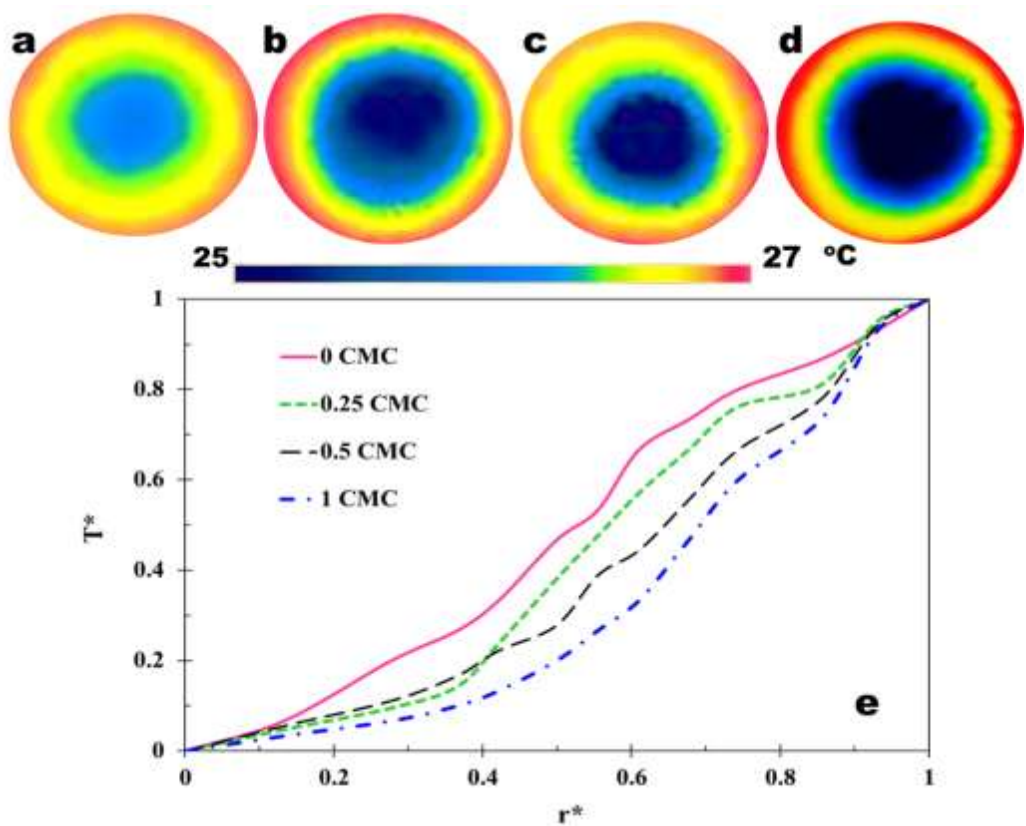

**Figure A4:** Infrared thermography images of the evaporating droplets on a SHS, for: a) 0 CMC, b) 0.25 CMC, c) 0.5 CMC, and d) 1 CMC concentrations. All the thermal contours here are top-view contours of the droplets. Here, the outer red-colored regions in the contours represent the substrate. The extracted non-dimensionalized temperature has been shown against the non-dimensional droplet radius in part e).

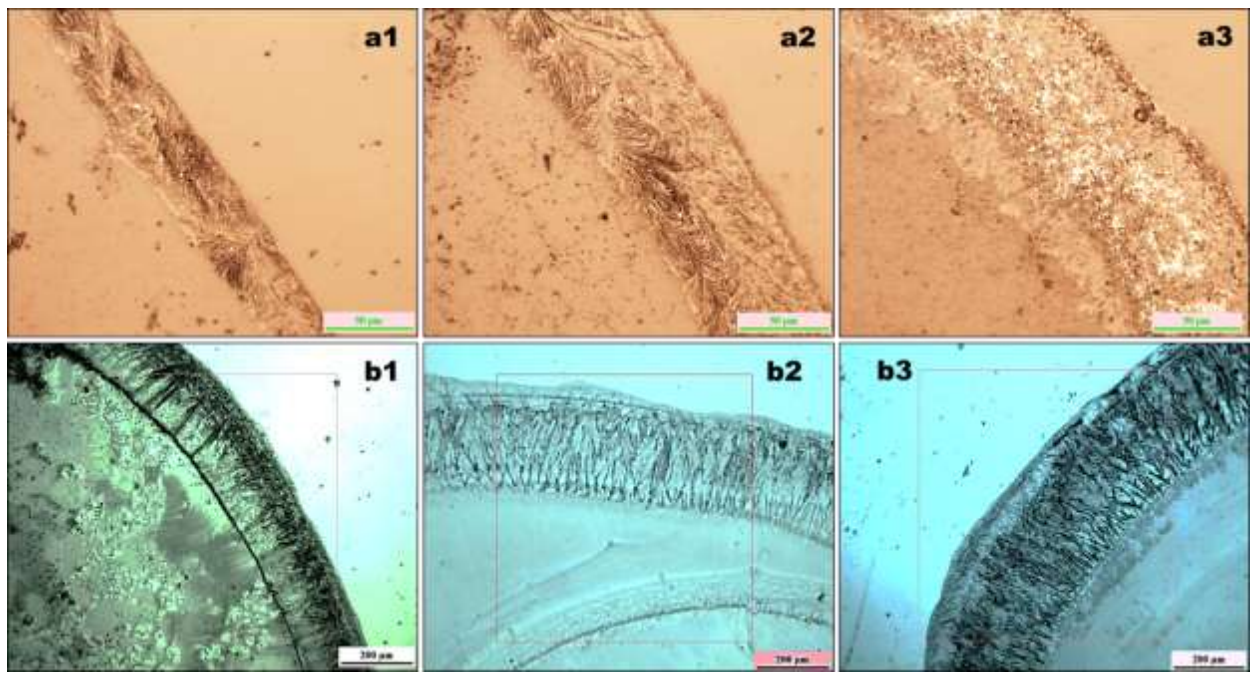

**Figure A5:** Formation of rims after drying for (a) Cetyltrimethylammonium bromide (CTAB) aqueous solution droplet at a1) 0.25 CMC, a2) 0.5 CMC, a3) CMC (50 µm scale), and (b) Sodium Dodecyl Sulphate (SDS) aqueous solution droplet at b1) 0.25 CMC, b2) 0.5 CMC, b3) 1 CMC (200 µm scale).

## REFERENCES

[1] J. Yin, Q. Di Wang, L.F. Zhang, L.K. Norvihoho, B. Liu, and Z.F. Zhou, "Evaporation characteristics of cis-1,1,1,4,4,4-hexafluoro-2-butene (R1336mzz(Z)) droplet in high pressure and temperature environments," Phys. Fluids **36**(2), (2024).

[2] J. Yin, B.J. Rong, Y. Liu, X.G. Zhu, Y.P. Li, Z.L. He, and Z.F. Zhou, "Comparative investigation on the droplet evaporation characteristics of hydrochlorofluorocarbon, hydrofluoroolefin, and hydrofluoroether under complex environments," Phys. Fluids **37**(7), (2025).

[3] S. Tonini, G.E. Cossali, E.A. Shchepakina, V.A. Sobolev, and S.S. Sazhin, "A model of droplet evaporation: New mathematical developments," Phys. Fluids **34**(7), (2022).

[4] V.N. Truskett, and K.J. Stebe, "Influence of surfactants on an evaporating drop: Fluorescence images and particle deposition patterns," Langmuir **19**(20), 8271–8279 (2003).

[5] V. Dugas, J. Broutin, and E. Souteyrand, "Droplet evaporation study applied to DNA chip manufacturing," Langmuir **21**(20), 9130–9136 (2005).

[6] S. Semenov, A. Trybala, H. Agogo, N. Kovalchuk, F. Ortega, R.G. Rubio, V.M. Starov, and M.G. Velarde, "Evaporation of droplets of surfactant solutions," Langmuir **29**(32), 10028–10036 (2013).

[7] F. He, Z. Wang, L. Wang, J. Li, and J. Wang, "Effects of surfactant on capillary evaporation process with thick films," Int. J. Heat Mass Transf. **88**, 406–410 (2015).

[8] P. Alexandridis, S.Z. Munshi, and Z. Gu, "Evaporation of water from structured surfactant solutions," Ind. Eng. Chem. Res. **50**(2), 580–589 (2011).

[9] V.I. Terekhov, and N.E. Shishkin, "Influence of a Surfactant on Evaporation Intensity of Suspended Water Droplets," Colloid J. **83**(1), 135–141 (2021).

[10] A. Marin, R. Liepelt, M. Rossi, and C.J. Kähler, "Surfactant-driven flow transitions in evaporating droplets," Soft Matter **12**(5), 1593–1600 (2016).

[11] J. Farhadi, and V. Bazargan, "Marangoni flow and surfactant transport in evaporating sessile droplets: A lattice Boltzmann study," Phys. Fluids **34**(3), (2022).

[12] N. Jung, H.W. Seo, P.H. Leo, J. Kim, P. Kim, and C.S. Yoo, "Surfactant effects on droplet dynamics and deposition patterns: A lattice gas model," Soft Matter **13**(37), 6529–6541 (2017).

[13] R.T. van Gaalen, C. Diddens, H.M.A. Wijshoff, and J.G.M. Kuerten, "Marangoni circulation in evaporating droplets in the presence of soluble surfactants," J. Colloid Interface Sci. **584**, 622–633 (2021).

[14] M. Wu, Y. Di, X. Man, and M. Doi, "Drying Droplets with Soluble Surfactants," Langmuir **35**(45), 14734–14741 (2019).

[15] R.T. van Gaalen, C. Diddens, H.M.A. Wijshoff, and J.G.M. Kuerten, "The evaporation of surfactant-laden droplets: A comparison between contact line models," J. Colloid Interface Sci. **579**, 888–897 (2020).

[16] X. Zhong, and F. Duan, "Dewetting transition induced by surfactants in sessile droplets at the early evaporation stage," Soft Matter **12**(2), 508–513 (2016).

[17] R. Bennacer, and X. Ma, "Effect of temperature and surfactants on evaporation and contact line dynamics of sessile drops," Heliyon **8**(11), e11716 (2022).

[18] A. Kaushal, V. Jaiswal, V. Mehandia, and P. Dhar, "Soluto-thermo-hydrodynamics influenced evaporation kinetics of saline sessile droplets," Eur. J. Mech. B/Fluids **83**, 130–140 (2020).

[19] A. Kaushal, V. Jaiswal, V. Mehandia, and P. Dhar, "Competing thermal and solutal advection decelerates droplet evaporation on heated surfaces," Eur. J. Mech. B/Fluids **91**, 129–140 (2022).

[20] A. Aldhaleai, and P.A. Tsai, "Evaporation dynamics of surfactant-laden droplets on a superhydrophobic surface: influence of surfactant concentration," Langmuir **38**(1), 593–601 (2021).

[21] A.R. Harikrishnan, P. Dhar, P.K. Agnihotri, S. Gedupudi, and S.K. Das, "Effects of interplay of nanoparticles, surfactants and base fluid on the surface tension of nanocolloids," Eur. Phys. J. E **40**(5), 16–23 (2017).

[22] A.R. Harikrishnan, and P. Dhar, "Optical thermogeneration induced enhanced evaporation kinetics in pendant nanofluid droplets," Int. J. Heat Mass Transf. **118**, 1169–1179 (2018).

[23] G. Karapetsas, K.C. Sahu, and O.K. Matar, "Evaporation of Sessile Droplets Laden with Particles and Insoluble Surfactants," Langmuir **32**(27), 6871–6881 (2016).

[24] P. Dhar, "Advection kinetics induced self-assembly of colloidal nanoflakes into microscale floral structures," Fluid Dyn. Res. **52**(6), (2020).

[25] M.J. Inanlu, B. Shojaan, J. Farhadi, and V. Bazargan, "Effect of particle concentration on surfactant-induced alteration of the contact line deposition in evaporating sessile droplets," Langmuir **37**(8), 2658–2666 (2021).

[26] A.R. Harikrishnan, P. Dhar, P.K. Agnihotri, S. Gedupudi, and S.K. Das, "Wettability of Complex Fluids and Surfactant Capped Nanoparticle-Induced Quasi-Universal Wetting Behavior," J. Phys. Chem. B **121**(24), 6081–6095 (2017).

[27] A.R. Harikrishnan, P. Dhar, S. Gedupudi, and S.K. Das, "Effect of Interaction of Nanoparticles and Surfactants on the Spreading Dynamics of Sessile Droplets," Langmuir **33**(43), 12180–12192 (2017).

[28] A.R. Harikrishnan, P. Dhar, P.K. Agnihotri, S. Gedupudi, and S.K. Das, "Correlating contact line capillarity and dynamic contact angle hysteresis in surfactant-nanoparticle based complex fluids," Phys. Fluids **30**(4), (2018).

[29] A.R. Harikrishnan, P. Dhar, S. Gedupudi, and S.K. Das, "Oscillatory solutothermal convection-driven evaporation kinetics in colloidal nanoparticle-surfactant complex fluid pendant droplets," Phys. Rev. Fluids **7**(3), 1–21 (2018).

[30] A.R. Harikrishnan, S.K. Das, P.K. Agnihotri, and P. Dhar, "Particle and surfactant interactions effected polar and dispersive components of interfacial energy in nanocolloids," J. Appl. Phys. **122**(5), (2017).

[31] A.R. Harikrishnan, P. Dhar, S. Gedupudi, and S.K. Das, "Governing influence of thermodynamic and chemical equilibria on the interfacial properties in complex fluids," J. Phys. Chem. B **122**(14), 4141–4148 (2018).

[32] K. Januszkiewicz, A. Mrozek-Niećko, and J. Różański, "Effect of surfactants and leaf surface morphology on the evaporation time and coverage area of ZnIDHA droplets," Plant Soil **434**(1–2), 93–105 (2019).

[33] Z. Zhou, C. Cao, L. Cao, L. Zheng, J. Xu, F. Li, and Q. Huang, "Effect of surfactant concentration on the evaporation of droplets on cotton (Gossypium hirsutum L.) leaves," Colloids Surfaces B Biointerfaces **167**, 206–212 (2018).

[34] Y. Li, V. Salvator, H. Wijshoff, M. Versluis, and D. Lohse, "Evaporation-Induced Crystallization of Surfactants in Sessile Multicomponent Droplets," Langmuir **36**(26), 7545–7552 (2020).

[35] K. Kotsi, T. Dong, T. Kobayashi, A. Moriarty, I. McRobbie, A. Striolo, and P. Angeli, "Effect of Surfactant Mixtures on the Evaporation Rate of Aqueous Sessile Droplets from Slightly Hydrophobic Substrates," Langmuir **41**(37), 25774–25788 (2025).

[36] A. Kaushal, V. Mehandia, and P. Dhar, "Ferrohydrodynamics governed evaporation phenomenology of sessile droplets," Phys. Fluids **33**(2), (2021).

[37] H. Hu, and R.G. Larson, "Analysis of the effects of marangoni stresses on the microflow in an evaporating sessile droplet," Langmuir **21**(9), 3972–3980 (2005).

[38] K. Sefiane, J.R. Moffat, O.K. Matar, and R. V. Craster, "Self-excited hydrothermal waves in evaporating sessile drops," Appl. Phys. Lett. **93**(7), 2006–2009 (2008).

[39] S. Dash, and S. V. Garimella, "Droplet evaporation dynamics on a superhydrophobic surface with negligible hysteresis," Langmuir **29**(34), 10785–10795 (2013).

[40] J.H. Snoeijer, and B. Andreotti, "Moving contact lines: Scales, regimes, and dynamical transitions," Annu. Rev. Fluid Mech. **45**, 269–292 (2013).

[41] R.D. Deegan, O. Bakajin, T.F. Dupont, G. Huber, S.R. Nagel, and T.A. Witten, "Capillaryflowasthecauseof ring stains fromdriedliquid drops," Nature **389**(6653), 827–829 (1997).

[42] Y.O. Popov, "Evaporative deposition patterns: Spatial dimensions of the deposit," Phys. Rev. E - Stat. Nonlinear, Soft Matter Phys. **71**(3), 1–17 (2005).

[43] E. Summerton, G. Zimbitas, M. Britton, and S. Bakalis, "Crystallisation of sodium dodecyl sulfate and the corresponding effect of 1-dodecanol addition," J. Cryst. Growth **455**(April), 111–116 (2016).

[44] C. Vautier-Giongo, and B.L. Bales, "Estimate of the ionization degree of ionic micelles based on Krafft temperature measurements," J. Phys. Chem. B **107**(23), 5398–5403 (2003).

[45] D.A. Nield, "Surface tension and buoyancy effects in cellular convection," J. Fluid Mech. **19**(3), 341–352 (1964).

[46] S.H. Davis, "Buoyancy-surface tension instability by the method of energy," J. Fluid Mech. **39**(2), 347–359 (1969).

[47] D.K. Mandal, and S. Bakshi, "Internal circulation in a single droplet evaporating in a closed chamber," Int. J. Multiph. Flow **42**, 42–51 (2012).

[48] T. Still, P.J. Yunker, and A.G. Yodh, "Surfactant-induced Marangoni eddies alter the coffee-rings of evaporating colloidal drops," Langmuir **28**(11), 4984–4988 (2012).

[49] C. Seo, D. Jang, J. Chae, and S. Shin, "Altering the coffee-ring effect by adding a surfactant-like viscous polymer solution," Sci. Rep. **7**(1), 1–9 (2017).

[50] T. Kajiya, W. Kobayashi, T. Okuzono, and M. Doi, "Controlling the drying and film formation processes of polymer solution droplets with addition of small amount of surfactants," J. Phys. Chem. B **113**(47), 15460–15466 (2009).

[51] W. Al-Soufi, L. Piñeiro, and M. Novo, "A model for monomer and micellar concentrations in surfactant solutions: Application to conductivity, NMR, diffusion, and surface tension data," J. Colloid Interface Sci. **370**(1), 102–110 (2012).

[52] R. De Dier, W. Sempels, J. Hofkens, and J. Vermant, "Thermocapillary fingering in surfactant-laden water droplets," Langmuir **30**(44), 13338–13344 (2014).

[53] A. Tiwari, S. Shin, S.J. Lee, and A.K. Thokchom, "Understanding the formation of particle bands and fingering patterns during evaporation of a sessile droplet containing colloids," Colloids Interface Sci. Commun. **56**(September), 100740 (2023).

[54] Á.G. Marín, H. Gelderblom, D. Lohse, and J.H. Snoeijer, "Order-to-disorder transition in ring-shaped colloidal stains," Phys. Rev. Lett. **107**(8), 1–4 (2011).

[55] X. Shao, F. Duan, Y. Hou, and X. Zhong, "Role of surfactant in controlling the deposition pattern of a particle-laden droplet: Fundamentals and strategies," Adv. Colloid Interface Sci. **275**, 102049 (2020).

[56] H. Li, H. Luo, Z. Zhang, Y. Li, B. Xiong, C. Qiao, X. Cao, T. Wang, Y. He, and G. Jing, "Direct observation of nanoparticle multiple-ring pattern formation during droplet evaporation with dark-field microscopy," Phys. Chem. Chem. Phys. **18**(18), 13018–13025 (2016).